\documentclass[iop,apj,revtex4-2,twocolappendix]{emulateapj}
\usepackage{graphicx}
\usepackage{amsmath,amssymb}

\begin{document}

\title{Spectral transfer and K\'arm\'an-Howarth-Monin equations for compressible Hall magnetohydrodynamics}

\author{Petr Hellinger\altaffilmark{1,2}}
\email{petr.hellinger@asu.cas.cz}
\author{Emanuele Papini\altaffilmark{3,4}}
\author{Andrea Verdini\altaffilmark{3,4}}
\author{Simone Landi\altaffilmark{3,4}}
\author{Luca Franci\altaffilmark{5}}
\author{Lorenzo Matteini\altaffilmark{6}}
\author{Victor Montagud-Camps\altaffilmark{1}}

\altaffiltext{1}{Astronomical Institute, CAS,
Bocni II/1401,CZ-14100 Prague, Czech Republic}
\altaffiltext{2}{Institute of Atmospheric Physics, CAS,
Bocni II/1401, CZ-14100 Prague, Czech Republic}
\altaffiltext{3}{Dipartimento di Fisica e Astronomia, Universit\`a degli Studi di Firenze Largo E. Fermi 2, I-50125 Firenze, Italy}
\altaffiltext{4}{INAF -- Osservatorio Astrofisico di Arcetri, Largo E. Fermi 5, I-50125 Firenze, Italy}
\altaffiltext{5}{Queen Mary University of London, UK}
\altaffiltext{6}{Imperial Colege, London, UK}

\date{\today}

\begin{abstract}
We derive two new forms of the
K\'arm\'an-Howarth-Monin equation for decaying compressible 
Hall magnetohydrodynamic (MHD) turbulence.
We test them on results of a weakly-compressible, two-dimensional,
moderate-Reynolds-number Hall MHD simulation
and compare them with  an isotropic spectral transfer (ST) equation.
The KHM and ST equations are automatically satisfied
during the whole simulation owing to the periodic
boundary conditions and have complementary 
cumulative behavior. They are used here to analyze the 
onset of turbulence and its properties when it is fully developed.
These approaches give equivalent results
characterizing: the decay of the kinetic + magnetic energy at large scales,
the MHD and Hall cross-scale energy transfer/cascade,
the pressure dilatation, and the dissipation. 
The Hall cascade appears when the MHD one brings
the energy close to the ion inertial range and is 
related to the formation of reconnecting current sheets. 
At later times, the pressure-dilation
energy-exchange rate oscillates around zero
with no net effect on the cross-scale energy transfer
when averaged over a period of its oscillations. 
A reduced one-dimensional analysis suggests that all
three methods may be useful to estimate
the energy cascade rate from in situ observations.
\end{abstract}



\maketitle

\section{Introduction}

Turbulence is ubiquitous in astrophysical plasma 
environments \citep{mave11}. The solar wind constitutes a natural laboratory for studying turbulence
in weakly collisional plasmas \citep{brca13}.
As it expands from the solar corona,
it exhibits a strong non-adiabatic behavior
and to sustain its thermal ion energetic properties it needs
to be heated \citep{vasqal07,cranal09,hellal11,hellal13}.
The situation is less clear for electrons
that carry a strong heat flux \citep{stveal15}.
The solar wind is a strongly turbulent flow
with large-amplitude fluctuations of the magnetic
field, particle velocity field, and other quantities as well.
These turbulent fluctuations are the usual suspect for 
the observed particle heating.

Besides phenomenological approaches, the K\'arm\'an-Howarth-Monin (KHM)
equation \citep{kaho38,kolm41b,moya75,fris95} represents a way how
to determine the cascade/dissipation rate in turbulence.
This equation was originally derived 
for incompressible (constant-density) hydrodynamic (HD) turbulence
 and further extended to incompressible magnetohydrodynamic 
(MHD) turbulence \citep{popo98b}
and to incompressible Hall MHD turbulence
\citep{galt08,hellal18,ferral19}.
Starting from the KHM equations and under further assumptions
(e.g., isotropy), the so called exact (scaling) laws can be derived
to connect the third-order structure functions with the energy cascade/dissipation rate.
In magnetized plasmas, the assumption of isotropy is, however,
questionable due to the anisotropy introduced by the ambient
magnetic field \citep{shebal83,oughal94}. Once 
relaxed, the cascade rate is given by the
divergence of third-order structure functions.

In situ spacecraft observations of turbulence
are based on one-dimensional time series
and it is not fully clear to what
extent these characterize
the inherently three-dimensional 
turbulent fluctuations, and how well the KHM equation can be used to estimate the
cascade rate \citep{podeal09,smital18}.
First applications of the KHM equation to 
in situ observations showed 
linear scaling of the third-order structure functions
\citep{sorral07,marial08}
and isotropy was assumed to estimate the corresponding
cascade rate. To account for the expected anisotropy,
a hybrid model that combines a two-dimensional (2D) perpendicular
(with respect to the ambient magnetic field)
cascade with a parallel one-dimensional (1D) was developed \citep{macbal08,stawal09}.
The turbulent energy cascade and its anisotropy can be partly constrained by
multi-spacecraft observations \citep{osmaal11},
but observational works need to be complemented by numerical simulations. 
\cite{verdal15} analyzed the effect of using
different turbulence models for estimating the
cascade rates using direct MHD simulations,
showing that the isotropic approximation
may lead to large errors in the estimation 
of the cascade rate. The hybrid model \citep{macbal08,stawal09}
gives, on the other hand, relatively good energy-cascade estimates.
In this respect,
\cite{franal20} compared in situ observations of turbulence 
with simulation results based on observation-driven plasma parameters,
finding good quantitative agreement between spectral properties
of the observed and simulated turbulent fluctuations, but also
a good agreement between the observed and simulated third-order structure functions
and the resulting cascade rates.

To date, there are not many studies estimating the energy cascade
rate in the solar wind and estimates based on the KHM equation
are typically sufficient to explain the observed temperature radial profiles
\citep{cobual15,smva21}, however, estimated cascade rates exhibit a large
variability and may even reach negative values in fast solar wind streams
with a large cross-helicity \citep{smital09}. 
Most of the works is done at 1 au (or further away from
the Sun. Recently, \cite{bandal20a} used data from the
Parker Solar Probe at 0.17 and 0.25 au and showed (assuming isotropy)
that the estimated cascade rate is consistent
with the radial profile of the proton temperature. However, a systematic
study of the radial dependence of the cascade rate is missing.
On the other hand, it is not clear, if solar wind turbulence
is strong enough to keep heating the plasma.
While the numerical study of \cite{montal18} shows that turbulence
is able sustain a radial profile of the temperature
similar to the observed one in the inner heliosphere
(at least within the MHD approximation with an isotropic temperature),
phenomenological transport models of solar wind turbulence
 \citep{zhma90,oughal11} indicate that, eventually at larger
radial distances from the Sun, turbulent fluctuations 
would be exhausted and need to be, in turn, sustained
by some energy injection.

The KHM equations can be also used to understand at which scales the dissipation
becomes important. Solar wind turbulence exhibits a transition in the form of 
a spectral break, at ion scales \citep{chenal14} and a similar transition is also observed
in direct numerical simulations \cite[e.g.,][and references therein]{franal16b,papial19}. 
Analyses of in situ observations and numerical simulations
 based on the incompressible Hall MHD KHM equation
suggest that this transition is a combination of
the onset of Hall physics and
dissipation \citep{hellal18,bandal20b}.
These results have some limitations:
the simulations are 2D, the observations are
based on 1D time series and assume isotropy.

The solar wind exhibits  weak density fluctuations,
$\delta n/n \sim 0.1$, so that the incompressible approximation
\cite[or the nearly-incompressible one, cf.,][]{zankal17}
is likely applicable.
On the other hand, closer to the Sun
the compressibility is expected to be larger 
\cite[cf.,][]{krupal20}.
Furthermore, at ion characteristic scales the level
of compressibility (density fluctuations) increases
\citep{chenal13b,pitnal19} so that the
incompressible approximation may not hold.
Also in high-cross helicity flows, the 
incompressible nonlinearity may be reduced 
and the compressible coupling may become important.
\cite{baneal16} and {hadial17} show that the compressible KHM 
equation gives larger cascade rates. 
\cite{andral19} show that the compressibility
may be important at sub-ion scales
where the Hall physics becomes important.
However, these results \cite[][and references therein]{baneal16,hadial17,andral19} have a major limitation,
they include the internal energy in the isothermal
closure (consequently they estimate the cascade
rate of the total energy including the redistribution
of the internal energy).

In order to avoid issues in the calculation of the divergence
of the third-order structure function,
\cite{baga17} proposed an alternative form of the
KHM equation (for the incompressible Hall MHD)
where the cascade rate is expressed using
second-order structure functions.
In this paper we analyze three different approaches
that may be used to estimate the cascade rates
of the kinetic + magnetic energy
in compressible Hall MHD turbulence. We derive a new
form of the compressible (standard) KHM equation 
\citep{hellal18,ferral19} and
we also generalize the alternative formulation of 
\cite{baga17} to the compressible Hall MHD.
We compare these KHM approaches with an isotropic spectral
transfer equation \citep{hellal21a}.
We use these methods to analyze results of
a 2D weakly compressible Hall MHD simulation.
We also look at reduced 1D results 
in the context of in situ observations.
The paper is organized as follows. 
In section~\ref{KHM}, we derive
the standard (subsection~\ref{standKHM}) and
the alternative KHM  (subsection~\ref{altKHM})
equations.
In section~\ref{simul}, we analyze the results
of a 2D Hall MHD simulation using the KHM equations.
In section \ref{spst}, we present and
apply the ST analysis to the Hall MHD simulation.
Results of the ST and KHM analyses are then compared.
In section~\ref{1Dred}, we analyze the energy cascade
rates from 1D reduced forms of the 
two KHM equation and of the ST one.
In section~\ref{discuss} we discuss the obtained
results.

\section{KHM equation}
\label{KHM}

We investigate a system governed by the following 
Hall MHD equations for the plasma density $\rho$,
the plasma mean velocity $\boldsymbol{u}$, and for the magnetic field
$\boldsymbol{B}$:
\begin{align}
\frac{\partial \rho}{\partial t}+(\boldsymbol{u}\cdot\boldsymbol{\nabla})\rho&=
-\rho \boldsymbol{\nabla}\cdot \boldsymbol{u}, \label{density}
\end{align}
\begin{align}
\rho\frac{\partial\boldsymbol{u}}{\partial t}+\rho(\boldsymbol{u}\cdot\boldsymbol{\nabla})\boldsymbol{u}&=
(\boldsymbol{\nabla}\times\boldsymbol{B})\times\boldsymbol{B}	
 -\boldsymbol{\nabla}p+ \boldsymbol{\nabla}\cdot\boldsymbol{\tau},
\label{HallMHDv}
\end{align}
\begin{align}
\frac{\partial\boldsymbol{B}}{\partial t}&=
\boldsymbol{\nabla}\times \left[(\boldsymbol{u}-\boldsymbol{j})\times\boldsymbol{B}\right]
+\eta\Delta\boldsymbol{B},
\label{HallMHD}
\end{align}
where $p$ is the plasma pressure,
$\boldsymbol{\tau}$
is the viscous stress tensor
(given by $\tau_{ij}=\mu\left(\partial u_{i}/\partial x_{j}+\partial u_{j}/\partial x_{i}-2/3\delta_{ij}\partial u_{k}/\partial x_{k}\right)$ where the dynamic viscosity $\mu$ is assumed to be constant),
 $\eta$ the electric resistivity, $\boldsymbol{j}$
is the electric current density in velocity units, $\boldsymbol{j}= \boldsymbol{J}/\rho_c=\boldsymbol{u}-\boldsymbol{u}_e$ ($\rho_c$ being the charge density). Here we assume SI units except for the magnetic permeability $\mu_0$ that is set to one
(SI results can be obtained by the rescaling $\boldsymbol{B} \rightarrow \boldsymbol{B} \mu_0^{-1/2}$).

For the formulation of the KHM equation in terms of structure functions,
we use the density-weighted velocity field $\boldsymbol{w}=\rho^{1/2} \boldsymbol{u}$
\citep{kior90,scgr19,prgi19} to take into account a variable density.
In order to represent
 the kinetic and magnetic energy we use the following structure functions
\begin{align*}
\mathcal{S}_{w} &=\left\langle |\delta\boldsymbol{w}|^2\right\rangle \ \ \ \text{and}  \ \ \   
\mathcal{S}_{B}=\left\langle |\delta\boldsymbol{B}|^{2}\right\rangle,
\end{align*}
respectively. Here $\delta\boldsymbol{w}=\boldsymbol{w}(\boldsymbol{x}^\prime)-\boldsymbol{w}(\boldsymbol{x})$, where
$\boldsymbol{x}^\prime=\boldsymbol{x}+\boldsymbol{l}$,
and $\langle \bullet \rangle$ denotes spatial averaging (over $\boldsymbol{x}$); the same
definition holds for $\delta\boldsymbol{B}$ and other quantities. Henceforth, we denote
the value of any quantity $a$ at $\boldsymbol{x}$ and $\boldsymbol{x}^\prime$
 as $a=a(\boldsymbol{x})$ and $a^\prime=a(\boldsymbol{x}^\prime)$, respectively.

\subsection{Standard KHM}
\label{standKHM}

For the total structure function, $\mathcal{S}=\mathcal{S}_{w}+\mathcal{S}_{B}$,
one can get a dynamic KHM equation, taking Eqs.~(\ref{HallMHDv}--\ref{HallMHD})
at two positions, $\boldsymbol{x}^\prime$ and $\boldsymbol{x}$.
Subtracting those
 one gets equations for $\partial \delta\boldsymbol{w}/\partial t$ and 
$\partial \delta\boldsymbol{B}/\partial t$.
Assuming a statistically homogeneous system one can get then the following form of the 
KHM equation (for the detailed derivation see Appendix)
\begin{align}
\frac{\partial \mathcal{S}}{\partial t}+
\boldsymbol{\nabla}_{\boldsymbol{l}}\cdot\left(\boldsymbol{\mathcal{Y}}+\boldsymbol{\mathcal{H}}\right)
 +\mathcal{R}&= \mathcal{C}
+2\left\langle \delta\theta\delta p\right\rangle -2\left\langle \delta\boldsymbol{\Sigma}:\delta\boldsymbol{\tau}\right\rangle  \nonumber \\
& -4 Q_\eta +2\eta\Delta_{\boldsymbol{l}} \mathcal{S}_{B},
\label{CYaglom}
\end{align}
Where $\theta=\boldsymbol{\nabla}\cdot \boldsymbol{u}$
is the dilatation field, $\boldsymbol{\Sigma}=\boldsymbol{\nabla}\boldsymbol{u}$ is the 
stress tensor, and $\Delta_{\boldsymbol{l}}$ denotes the Laplace operator (with
respect to the separation $\boldsymbol{l}$).
Eq.~(\ref{CYaglom}) combines the second-order structure functions 
with the third-order ones related to the energy transfer/cascade rate
\begin{align*}
\boldsymbol{\mathcal{Y}}&=\left\langle \delta\boldsymbol{u} \left(|\delta\boldsymbol{w}|^2+
 |\delta\boldsymbol{B}|^{2}\right)-2\delta\boldsymbol{B}\left(\delta\boldsymbol{u}\cdot\delta\boldsymbol{B}\right)\right\rangle \nonumber \\
\boldsymbol{\mathcal{H}}&= \left\langle \delta\boldsymbol{B}\left(\delta\boldsymbol{j}\cdot\delta\boldsymbol{B}\right)-\frac{1}{2}\delta\boldsymbol{j}|\delta\boldsymbol{B}|^{2}\right\rangle.
\end{align*}
Eq.~(\ref{CYaglom}) also contains 
the explicitly compressible contribution to the cascade rate, 
\begin{align*}
\mathcal{R}=\left\langle 
\delta\boldsymbol{w}\cdot \left( \theta^\prime \boldsymbol{w}-
\theta \boldsymbol{w}^\prime \right)\right\rangle,
\end{align*} 
 of hydrodynamic origin \cite[cf.,][]{hellal21a}.
The dissipation part includes the resistive incompressible-like terms
$Q_{\eta}- \eta \Delta_{\boldsymbol{l}} \mathcal{S}_B/2$
and the generally compressible viscous term
$\left\langle \delta\boldsymbol{\tau}:\delta\boldsymbol{\Sigma}\right\rangle/2$,
where $Q_{\eta} = \eta\left\langle \boldsymbol{\nabla} \boldsymbol{B} :
\boldsymbol{\nabla} \boldsymbol{B}\right\rangle
= \eta \left\langle |\boldsymbol{J}|^2\right\rangle$
is the Joule 
dissipation rate per unit volume ($\boldsymbol{J}=\boldsymbol{\nabla}\times \boldsymbol{B}$ being the electric current density).
The correction term $\mathcal{C}$ can be given in the form
\begin{align*}
\mathcal{C}
&=2C_{\sqrt{\rho}}\left[\boldsymbol{u},\boldsymbol{\nabla} p \right]
-2C_{\sqrt{\rho}}\left[\boldsymbol{u},\boldsymbol{\nabla}\cdot\boldsymbol{\tau}\right] \nonumber \\
&+ 2 C_{\sqrt{\rho}}\left[\boldsymbol{u},\boldsymbol{B}\times\boldsymbol{J}\right]
+C_{\rho}\left[\boldsymbol{B}\times\boldsymbol{j},\boldsymbol{J}\right],
\end{align*}
where
\begin{align*}
C_{\rho}\left[\boldsymbol{a},\boldsymbol{b}\right]
=\left\langle\left(\frac{\rho^{\prime}}{\rho}-1\right)\boldsymbol{a}^{\prime}\cdot\boldsymbol{b}+\left(\frac{\rho}{\rho^{\prime}}-1\right)\boldsymbol{a}\cdot\boldsymbol{b}^{\prime} \right\rangle.
\end{align*}
Note that this term explicitly depends on the density variation
and disappear in the constant density approximation.

Eq.~(\ref{CYaglom}) can be cast in the following form
\begin{align}
\frac{1}{4}\frac{\partial \mathcal{S}}{\partial t}
-\mathcal{K}_\text{H} -\mathcal{K}_\text{MHD} = \varPsi-\mathcal{D}
\label{CYaglomb}
\end{align}
where we have combined some terms as
\begin{align}
\mathcal{K}_\text{MHD}&=-\frac{1}{4}\boldsymbol{\nabla}\cdot\boldsymbol{\mathcal{Y}} -\frac{1}{4} \mathcal{R} \nonumber
+\frac{1}{2} C_{\sqrt{\rho}}\left[\boldsymbol{u},\boldsymbol{B}\times\boldsymbol{J}\right] \\
\mathcal{K}_\text{H}&=-\frac{1}{4}\boldsymbol{\nabla}\cdot\boldsymbol{\mathcal{H}}
+\frac{1}{4} C_{\rho}\left[\boldsymbol{B}\times\boldsymbol{j},\boldsymbol{J}\right] \\
\varPsi &= \frac{1}{2} \langle \delta p \delta \theta \rangle  + \frac{1}{2}  C_{\sqrt{\rho}}\left[\boldsymbol{u},\boldsymbol{\nabla} p \right] \nonumber
\\
\mathcal{D}&=Q_{\eta}- \frac{\eta}{2} \Delta \mathcal{S}_B
+ \frac{1}{2} \left\langle \delta\boldsymbol{\tau}:\delta\boldsymbol{\Sigma}\right\rangle
+ \frac{1}{2} C_{\sqrt{\rho}}\left[\boldsymbol{u},\boldsymbol{\nabla}\cdot\boldsymbol{\tau}\right]\nonumber,
\end{align}
and where we dropped the ${\boldsymbol{l}}$ index for $\boldsymbol{\nabla}$ and $\Delta$. 
Here, $\mathcal{K}_\text{MHD}$ and $\mathcal{K}_\text{H}$ are the MHD and Hall
cascade rates, respectively, $\varPsi$ represents the pressure-dilatation effect whereas
$\mathcal{D}$ accounts for the effects of dissipation and heating.

As noted by \cite{hellal21a}, the KHM structure functions in hydrodynamic (HD) turbulence
have a cumulative behavior that is complementary to the isotropic ST equation.
We will see in section~\ref{compKHMST} that this is also true in
Hall MHD. Here we just note that
the viscous term 
$\left\langle \delta\boldsymbol{\tau}:\delta\boldsymbol{\Sigma}\right\rangle$ 
 is the generalization to the compressible case of the two incompressible dissipative terms
$2 \nu \left\langle \boldsymbol{\nabla} \boldsymbol{u} :
\boldsymbol{\nabla} \boldsymbol{u}\right\rangle - \nu \Delta
\left\langle \delta \boldsymbol{u}^2 \right\rangle$.
The behavior of $\left\langle \delta\boldsymbol{\tau}:\delta\boldsymbol{\Sigma}\right\rangle$
changes with scales.
At large scales $|\boldsymbol{l}|\rightarrow \infty$, where the correlations 
$\left\langle {\boldsymbol{\tau}(\boldsymbol{x}^\prime)}:\boldsymbol{\Sigma} 
(\boldsymbol{x})\right\rangle\rightarrow 0$
the viscous term becomes twice the viscous heating rate
$Q_{\mu}$,
$\left\langle \delta\boldsymbol{\tau}:\delta\boldsymbol{\Sigma}\right\rangle  \rightarrow 2 \left\langle \boldsymbol{\tau}:\boldsymbol{\Sigma}\right\rangle = 2 Q_{\mu}$
and the dissipation term reaches 
$\mathcal{D} \rightarrow Q_{\text{tot}}$,
where we denote the total heating rate as $Q_{\text{tot}}=Q_\mu+Q_\eta$.

\subsection{Alternative formulation}

\label{altKHM}

The cross-scale energy transfer/cascade rates
in Equation~\ref{CYaglom} contains terms
with a divergence of third-order structure functions
that are difficult to evaluate from 1D time series.
\cite{baga17} presented an alternative
formulation of the KHM equation for incompressible Hall MHD turbulence
where these terms are replaced by second-order
structure functions. This approach 
 can be easily generalized
to the compressible case and the resulting
equation can be given the following form

\begin{align}
\frac{\partial \mathcal{S}}{\partial t}+
\tilde{\mathcal{Y}}+\tilde{\mathcal{H}}
 +\tilde{\mathcal{R}}&= \tilde{\mathcal{C}}
+2\left\langle \delta\theta\delta p\right\rangle -2\left\langle \delta\boldsymbol{\Sigma}:\delta\boldsymbol{\tau}\right\rangle  \nonumber \\
& -4 Q_\eta +2\eta\Delta \mathcal{S}_{B}
\label{CYaglom2}
\end{align}
where
\begin{align}
\tilde{\mathcal{Y}} &= -2\left\langle \delta\boldsymbol{w}\cdot\delta\left(\boldsymbol{u}\times \boldsymbol{\varpi}\right)\right\rangle
-2\left\langle \delta\boldsymbol{w}\cdot\delta\left(\frac{\boldsymbol{J}\times\boldsymbol{B}}{\sqrt{\rho}}\right)\right\rangle \nonumber \\
&-2\left\langle \delta\boldsymbol{J}\times\delta\left(\boldsymbol{u}\times\boldsymbol{B}\right)\right\rangle   \\
\tilde{\mathcal{H}} &= 2\left\langle \delta\boldsymbol{J}\times\delta\left(\boldsymbol{j}\times\boldsymbol{B}\right)\right\rangle  \\
\tilde{\mathcal{R}} &=\left\langle \delta\boldsymbol{w}\cdot\delta\left(\boldsymbol{w}\theta\right)\right\rangle 
+2\left\langle \delta\boldsymbol{w}\cdot\delta\left[(\boldsymbol{\nabla}\boldsymbol{w})\cdot\boldsymbol{u}\right]\right\rangle  \\
\tilde{\mathcal{C}}
&=2C_{\sqrt{\rho}}\left[\boldsymbol{u},\boldsymbol{\nabla} p \right]
-2C_{\sqrt{\rho}}\left[\boldsymbol{u},\boldsymbol{\nabla}\cdot\boldsymbol{\tau}\right] 
\end{align}
and $\boldsymbol{\varpi}=\boldsymbol{\nabla}\times\boldsymbol{w}$. The other terms are defined in
the previous section.

From Eq.(\ref{CYaglom2}) we get the MHD and  Hall cascade rates as
\begin{align}
\mathcal{K}_\text{MHD}= -\frac{1}{4}\tilde{\mathcal{Y}} - \frac{1}{4} \tilde{\mathcal{R}}, \ \ \
\mathcal{K}_\text{H}= -\frac{1}{4} \tilde{\mathcal{H}}.
\end{align}
Note that only the first term in $\tilde{\mathcal{R}}$ disappears in the
incompressible, constant-density limit. The second term,
in this limit, is proportional to $\langle \delta \boldsymbol{u} \cdot
\delta( \boldsymbol{\nabla}|\boldsymbol{u}|^2) \rangle$
\cite[cf.,][]{baga17} and is generally non-negligible.

\section{Numerical simulation}
\label{simul}
Now we test the predictions of the compressible KHM equations
using a Hall MHD simulation.
The Hall MHD model consists of the nonlinear, compressible
viscous-resistive MHD equations, modified only by the presence of the 
Hall term in the induction equation, The system is
described by Eqs.~(\ref{density}--\ref{HallMHD}),
complemented by the equation for the plasma temperature 
\begin{align}
\frac{\partial T}{\partial t}+(\boldsymbol{u}\cdot\boldsymbol{\nabla})T&=
(\gamma-1)\left(
-T \boldsymbol{\nabla}\cdot \boldsymbol{u}+\frac{\eta|\boldsymbol{J}|^2}{\rho}
+ \frac{\boldsymbol{\tau}: \boldsymbol{\Sigma}}{\rho}\right),
\label{temperature}
\end{align}
where $\gamma=5/3$.
These equations are solved 
using the pseudo-spectral approach and a 3rd-order Runge-Kutta scheme \citep{papial19,papial19b}.
We consider a 2D $(x,y)$ periodic domain and use Fourier decomposition to calculate the spatial derivatives. In the Fourier space we also filter according to the $2/3$ Orszag rule \citep{orsz71}, 
to avoid aliasing of the quadratic nonlinear terms.
Aliasing of the cubic terms is mitigated by the
presence of a finite dissipation \citep{ghosal93}.
We consider a 2D box of size $256 d_i \times 256 d_i$ and a grid resolution of $\Delta x = \Delta y = d_i/8$, corresponding to $2048^2$ points, where $d_i$ is the ion inertial length.
We set a constant background magnetic field $\boldsymbol{B}_{0}$ along the $z$ (out-of-plane) direction.
The initial state is populated by large-amplitude Alfv\'enic fluctuations in the $xy$-plane up to the injection scale $\ell_\text{inj}=2\pi d_i/k_\perp^\text{inj}$, with $k_\perp^\text{inj} d_i \simeq 0.2$, where  $k_\perp = \sqrt{k_x^2+k_y^2}$.
The relative root-mean-square (rms) amplitude of these fluctuations is set to $0.17$
in units of $B_0=|\boldsymbol{B}_{0}|$. 
The plasma beta is initially $\beta=0.5$.
 No forcing is present, so the simulation resolves the evolution of
 freely decaying turbulence and we assume the viscosity and resistivity $\mu=\eta= 10^{-3}$.

\begin{figure}
\centerline{\includegraphics[width=8cm]{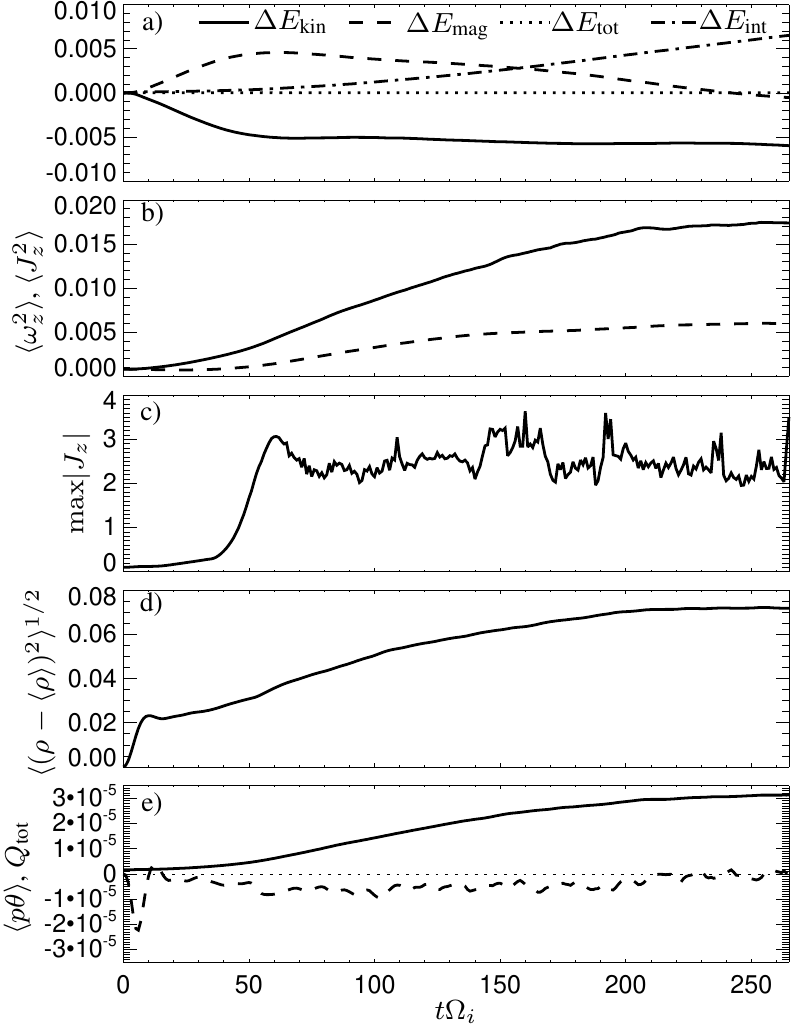}}
\caption{Evolution of different quantities as a function of time: (a) the relative changes
in the kinetic energy $\Delta E_{\text{kin}}$ (solid line), the magnetic energy  $\Delta E_{\text{mag}}$
(dashed line), the total energy
 $\Delta E_{\text{tot}}$ (dotted line), and the internal energy $\Delta E_{\text{int}}$ (dash-dotted),
 (b) squared rms values of the out-of-plane components of 
the current $J_z$ and the vorticity $\omega_z$ (dashed line), 
(c) the maximum value of $J_z$,
 (d) the rms value of the density fluctuations, and
(e) the total dissipation rate $Q_{\text{tot}}$ (solid line) and the pressure-dilatation
 rate $\langle p \theta \rangle$ (dashed line);
the dotted line denotes the zero level for comparison.
\label{evol}
}
\end{figure}

In the simulation the total energy $E_{\text{tot}}=E_{\text{kin}}+E_{\text{int}}+E_{\text{mag}}$
 is well conserved. Here $E_{\text{kin}}=\langle \rho u^2 \rangle/2$
is the kinetic energy, $E_{\text{int}}=\langle \rho T \rangle/(\gamma-1)$ is
the internal one, $E_{\text{mag}}=\langle B^2 \rangle /2$ 
(here $\langle \bullet \rangle$ denotes spatial averaging over the simulation box).
Fig.~\ref{evol}a
displays the evolution of the relative changes in these energies,
$\Delta E(t)=[E(t)-E(0)]/E_{\text{tot}}(0)$, where
$E=E_{\text{kin},\text{tot},\text{mag},\text{int}}$.  
The relative 
change 
of the total energy is negligible, $\Delta E_{\text{tot}}(t=8)\sim 3\ 10^{-5}$.
Fig.~\ref{evol}b shows the evolution
of the rms values of the out-of-plane components of
the current $J_z$ and the vorticity $\omega_z$.
$\langle |J_z|^2 \rangle $ reaches a maximum at $t\simeq 255 \Omega_i^{-1}$, which is a signature
of a fully developed turbulent cascade \citep{mipo09}. The $z$ component of
the vorticity exhibits a similar evolution.
Fig.~\ref{evol}c  shows the maximum value (over the simulation box) of $J_z$.
The sharp increase of $\mathrm{max}|J_z|$ for $ 40  \lesssim t \Omega_i \lesssim 60$
indicates the formation of thin current sheets and the saturation at $t \Omega_i \sim 60$
is due to the onset of magnetic reconnection \citep{franal17,papial19}.
Fig.~\ref{evol}d displays the rms value of the density, $\langle (\rho - \langle\rho\rangle)^2  \rangle^{1/2}$, which steadily increases until at $t \gtrsim 210  \Omega_i^{-1}$
there is an indication of a saturation.
Fig.~\ref{evol}e shows the dissipation and the pressure-dilatation rates.
The dissipation rate increases with time and becomes quasi-stationary
at later times. The pressure-dilatation rate becomes initially negative 
but at later times, $t \gtrsim 210  \Omega_i^{-1}$, it oscillates around zero.

\begin{figure}[htb]
\centerline{
\includegraphics[width=8cm]{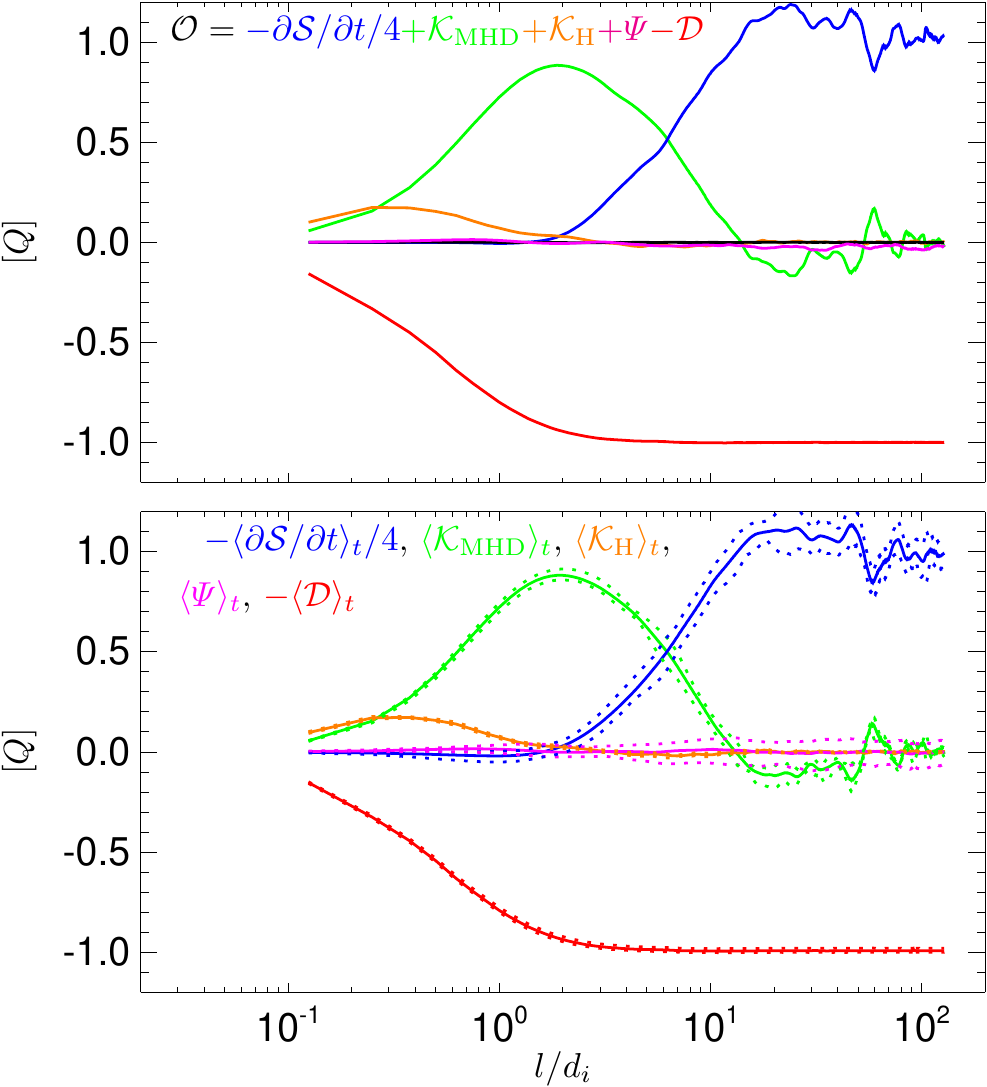}
}
\caption{KHM analysis: (top) The validity test of the compressible
KHM $\mathcal{O}$, Eq.~(\ref{oo}) (black line), as a function of $l$ at $t=255\Omega_i^{-1}$
along with the different
contributing terms: 
(blue) the decay $-{\partial \mathcal{S}}/{\partial t}/4$, 
(green) the MHD cascade $\mathcal{K}_\text{MHD}$,
(orange) the Hall cascade  $\mathcal{K}_\text{H}$, 
(red) the dissipation $-\mathcal{D}$, and 
(magenta) the pressure-dilatation $\varPsi$.
(bottom) Time-averaged ($240 \le t \Omega_i \le  264 $) contributing terms (solid lines)
with their minimum and maximum values (dotted lines) for
(blue) the decay $-\langle {\partial \mathcal{S}}/{\partial t}\rangle_t/4$,
(green) the MHD cascade $\langle \mathcal{K}_\text{MHD}\rangle_t$,
(orange) the Hall cascade  $\langle\mathcal{K}_\text{H}\rangle_t$,
(red) the dissipation/heating term $-\langle\mathcal{D}\rangle_t$, and
(magenta) the pressure-dilatation $\langle\varPsi\rangle_t$.
All the terms are normalized with respect to the total heating rate
$Q$.
\label{yaghall}}
\end{figure}

To check the prediction of Eq.~(\ref{CYaglomb}) in the simulation we define the validity test as
\begin{align}
\mathcal{O}= -\frac{1}{4}\frac{\partial \mathcal{S}}{\partial t}
+\mathcal{K}_\text{H}+\mathcal{K}_\text{MHD}
+ \varPsi -\mathcal{D}.
\label{oo}
\end{align}
The top panel of Figure~\ref{yaghall} shows this validity check $\mathcal{O}$
in the 2D simulation at $t=255 \Omega_i^{-1}$
as well as the contributing terms.
The compressible KHM equation is very well conserved:
$|\mathcal{O}|/Q <  0.3~\%$, where $Q$ is the heating/dissipation rate
at $t=255 \Omega_i^{-1}$, $Q=Q_{\text{tot}}(255 \Omega_i^{-1})$;
henceforth we use $Q$ to normalize all the relevant
quantities.
At large scales, $\partial \mathcal{S}/\partial t/4 -\varPsi\sim -\mathcal{D} \sim -Q$, that represents the energy conservation. 
The pressure-dilatation term $\varPsi$ is small.
At intermediate scales the MHD cascade dominates, $\mathcal{K}_\text{MHD}
\sim 0.85 Q$. This term is mostly compensated by the dissipation
term $\mathcal{D}$. At small scales the Hall cascade sets in
and becomes comparable to the MHD one. One these scales the
viscous and resistive dissipation also starts to act.

The pressure-dilatation effect is weak but not negligible.
Since the energy exchanges owing to this effect oscillate 
around zero, it is interesting to look at the average
behavior of the KHM equation over one period of these oscillations
\cite[see Fig.~\ref{evol}; cf.,][]{hellal21a}.
The bottom panel of Figure~\ref{yaghall} displays 
the different terms contributing to the
KHM equation averaged over such a period ($240 \le t \Omega_i 
 \le  264 $) as well as their minimum and maximum values
over this time (here $\langle \bullet \rangle_t$ denotes the
time average) .
The pressure-dilatation term varies within few percents 
of $Q$ over this period, and, on average, the effect
of the pressure-dilatation effect is negligible. 
The terms $-{\partial \mathcal{S}}/{\partial t}/4$
and $\mathcal{K}_\text{MHD}$ exhibit similar variability
whereas $\mathcal{K}_\text{H}$ and $\mathcal{D}$
are about constant.

\subsection{Evolution}

\begin{figure}
\centerline{\includegraphics[width=8cm]{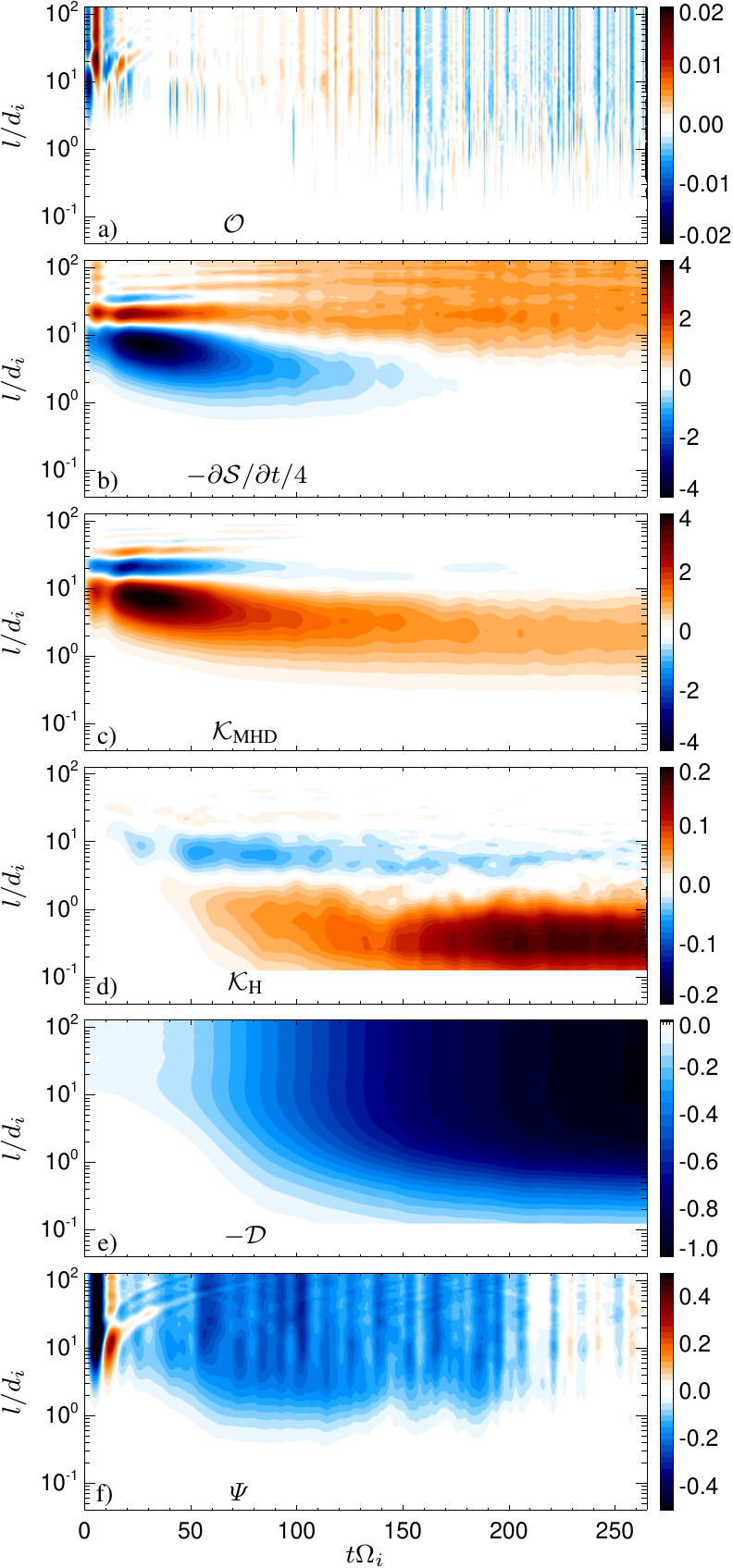}}
\caption{Evolution of the compressible KHM law:
color scale plots of 
(a) $\mathcal{O}$, (b) $-\partial \mathcal{S}/\partial t/4$, (c)
$\mathcal{K}_\text{MHD}$, (d) $\mathcal{K}_\text{H}$, (e) $\mathcal{D}$, and
(f) $\varPsi$ as functions of time and scale $l$.
All the terms are normalized with respect to the total heating rate
$Q$.
\label{yevol}
}
\end{figure}

The compressible KHM equation is valid during the whole simulation,
the homogeneity
condition, $\langle(\boldsymbol{\nabla}_{\boldsymbol{x}^\prime}+\boldsymbol{\nabla}_{\boldsymbol{x}}) \bullet \rangle=0$,
is automatically satisfied for the periodic boundary conditions.
Figure~\ref{yevol} displays the validity test
$\mathcal{O}$ as well as the different contributing
terms (normalized to $Q$)  as functions of time and $l$.
Figure~\ref{yevol}a shows that
the compressible KHM equation is well conserved,
$\mathrm{max}|\mathcal{O}|/Q \lesssim 1 \%$ for $t > 20 \Omega_i^{-1}$.
During early times, the maximum relative error in $|\mathcal{O}|/Q$ is about $2 \%$.
At later times, the $|\mathcal{O}|/Q$ is below $1 \%$ and is
fluctuating with time and roughly constant at all scales at a given time. This is
caused by numerical errors; in particular,
the time derivative $\partial \mathcal{S}/\partial t$ is estimated
by the finite difference with $\Delta t=  \Omega_i^{-1}$.
This is likely not sufficient for the transition from
the initial superposition of large scale Alfv\'enic
fluctuations. The interpretation is supported by the fact that $\mathcal{O}$ increases
when we increase $\Delta t$.

Figure~\ref{yevol}bc show that
initially $\partial\mathcal{S}/\partial t$ and $\mathcal{K}_\text{MHD}$
compensate each other, $\partial \mathcal{S}/\partial t + 4 \mathcal{K}_\text{MHD} \sim 0$.
$\partial \mathcal{S}/\partial t$ evolves and, at later times, dominates at large
scales indicating a decay of the kinetic + magnetic energy at large scales.
$\mathcal{K}_\text{MHD}$ becomes dominant at intermediate scales.
Figure~\ref{yevol}d shows that the Hall term becomes important
at $ 40  \lesssim t \Omega_i \lesssim 60$ for $l \lesssim 3 d_i$; this is about
the time when thin current sheets form and start to reconnect \citep{papial19}.
Interestingly, there is an indication of a negative Hall cascade
rate,
starting earlier  $ 40  \lesssim t \Omega_i$  for  $ 3 d_i \lesssim l \lesssim 20$,
 indicating an energy transfer from small to large scales, that
may be part of the onset of turbulence \cite[cf.,][]{franal17}.
The Hall term 
settles to the asymptotic value after $t\simeq 150$, when the energy at 
large scales has had time to participate to the cascade.

Figure~\ref{yevol}e demonstrates that the dissipation gradually develops
 as the (MHD and Hall) cascade evolves and brings energy to small scales.
Figure~\ref{yevol}f displays the pressure-dilatation term $\varPsi$.
After the initial transition, this term is important
on relatively large scales for a long time; its negative
value indicates a compressible plasma heating. Then it becomes
weakly oscillating around zero for $t \gtrsim 210 \Omega_i^{-1}$.

Note that the alternative KHM equation, Eq.~(\ref{CYaglom2}), 
gives, unsurprisingly, almost the same results as Eq.~(\ref{CYaglom}). 
The cascade rates $\mathcal{K}_\text{MHD}$ and $\mathcal{K}_\text{H}$ differ within $10^{-3}Q$
between the two methods.
On the other hand, for the weakly-compressible simulation,
 the incompressible approximation \citep{hellal18,ferral19}
exhibits an error that is mostly related to the neglected pressure-dilatation term.
However, the incompressible MHD and Hall cascade rates are close to their
compressible counterparts (not shown).

\section{Spectral Transfer}
\label{spst}

Another way to analyze the scale dependence 
of turbulence and its processes is the spectral (Fourier) decomposition.
To characterize the kinetic energy 
in the compressible flow one can use 
the density-weighted velocity field $\boldsymbol{w}$
as in the KHM approach.
The evolution for the energy in a given Fourier mode 
\begin{align}
\widehat{\boldsymbol{w}} (\boldsymbol{k})&= \sum_{\boldsymbol{x}} \boldsymbol{w}(\boldsymbol{x})
 \mathrm{exp}(i \boldsymbol{k} \cdot \boldsymbol{x} ),  \\
 \widehat{\boldsymbol{B}} (\boldsymbol{k})&= \sum_{\boldsymbol{x}} \boldsymbol{B}(\boldsymbol{x})
 \mathrm{exp}(i \boldsymbol{k} \cdot \boldsymbol{x} )
\end{align}
follows from Eqs.~(\ref{density}--\ref{HallMHD}) \cite[cf.,][]{minial07,gretal17}
\begin{align}
\frac{1}{2}\frac{\partial \left(|\widehat{\boldsymbol{w}}|^{2}+|\widehat{\boldsymbol{B}}|^{2}\right)}{\partial t}=& -T_{\text{MHD}}-T_{\text{H}}
-\Re\widehat{\boldsymbol{w}}^{*}\cdot\widehat{\frac{\boldsymbol{\nabla}p}{\sqrt{\rho}}} \nonumber \\ 
&+\Re\widehat{\boldsymbol{w}}^{*}\cdot\widehat{\frac{\boldsymbol{\nabla}\cdot\boldsymbol{\tau}}{\sqrt{\rho}}}
-\eta k^{2}|\widehat{\boldsymbol{B}}|^{2},
\label{wbk}
\end{align}
where the MHD and Hall transfer terms are 
\begin{align*}
T_{\text{MHD}}=& \Re \bigg[ \widehat{\boldsymbol{w}}^{*}\cdot\widehat{(\boldsymbol{u}\cdot\boldsymbol{\nabla})\boldsymbol{w}}+\frac{1}{2}\widehat{\boldsymbol{w}}^{*}\cdot\widehat{\boldsymbol{w}\theta} \nonumber \\
& -\widehat{\boldsymbol{w}}^{*}\cdot\widehat{\frac{(\boldsymbol{\nabla}\times\boldsymbol{B})\times\boldsymbol{B}}{\sqrt{\rho}}} 
 -\widehat{\boldsymbol{B}}^{*}\cdot\widehat{\boldsymbol{\nabla}\times\left(\boldsymbol{u}\times\boldsymbol{B}\right)} \bigg],
\\
T_{\text{H}} =&\Re \bigg[ \widehat{\boldsymbol{B}}^{*}\cdot\widehat{\boldsymbol{\nabla}\times\left(\boldsymbol{j}\times\boldsymbol{B}\right)} \bigg],
\end{align*}
respectively; here wide hat denotes the Fourier transform, the star denotes the complex
conjugate, and $\Re$ denotes the real part.

In the inertial range, one expects that the time-derivative terms in Eq.~(\ref{wbk}) are zero and 
the same is expected for the dissipation and pressure-dilatation terms. The
transfer term $T_{\text{MHD}}+T_{\text{H}}$ is also expected to be zero, as
all the energy that arrives from larger scales proceeds to smaller scales.
To characterize (isotropic) turbulence we use a
 low-pass filtered kinetic + magnetic energy 
\cite[i.e., the energy in modes with wave-vector magnitudes smaller than
or equal to $k$, cf.,][]{hellal21a}
\begin{align}
E_{k}&=\frac{1}{2}\sum_{|\boldsymbol{k}^\prime|\le k}\left(|\widehat{\boldsymbol{w}}|^{2}+|\widehat{\boldsymbol{B}}|^{2}\right).
\end{align}
For $E_{k}$ one 
 gets the following dynamic equation
\begin{align}
\frac{\partial E_{k}}{\partial t}+S_{\text{MHD}k}+S_{\text{H}k}=\Psi_{k}-D_{k}
\label{sptrdyn}
\end{align}
where 
\begin{align}
S_{\text{MHD}k}	&=\sum_{|\boldsymbol{k}^\prime|\le k} T_{\text{MHD}}(\boldsymbol{k}^\prime), \ \ \
S_{\text{H}k}	=\sum_{|\boldsymbol{k}^\prime|\le k} T_{\text{H}}(\boldsymbol{k}^\prime),
\\
\Psi_{k}&=-\sum_{|\boldsymbol{k}^\prime|\le k} \widehat{\boldsymbol{w}}^{*}\cdot\widehat{\rho^{-1/2}\boldsymbol{\nabla}p},
\\
D_{k}&=\eta \sum_{|\boldsymbol{k}^\prime|\le k}  {k^\prime}^{2}|\widehat{\boldsymbol{B}}|^{2}-\sum_{|\boldsymbol{k}^\prime|\le k} \widehat{\boldsymbol{w}}^{*}\cdot\widehat{\rho^{-1/2}\boldsymbol{\nabla}\cdot\boldsymbol{\tau}}.
\end{align}
Here $S_{\text{MHD}k}$ and $S_{\text{H}k}$ represent the MHD and Hall energy transfer rates, respectively, 
$\Psi_k$ describes the pressure-dilatation effect,
and $D_k$ is the (viscous and resistive) dissipation rate for modes with wave-vector magnitude smaller than or equal to $k$.
When $S_{\text{MHD}k}$ or $S_{\text{H}k}$ are constant,
the corresponding energy transfer rate is constant,
and can be identified with the cascade rate.
For large wave vectors, one gets the unfiltered values
\begin{align}
E_{k}\rightarrow E_{\text{kin}}+E_{\text{mag}}, \ \ \
\Psi_k \rightarrow \langle p\theta \rangle, \ \ \text{and}   \  \
D_k  \rightarrow Q,
\end{align}

\begin{figure}[htb]
\centerline{
\includegraphics[width=8cm]{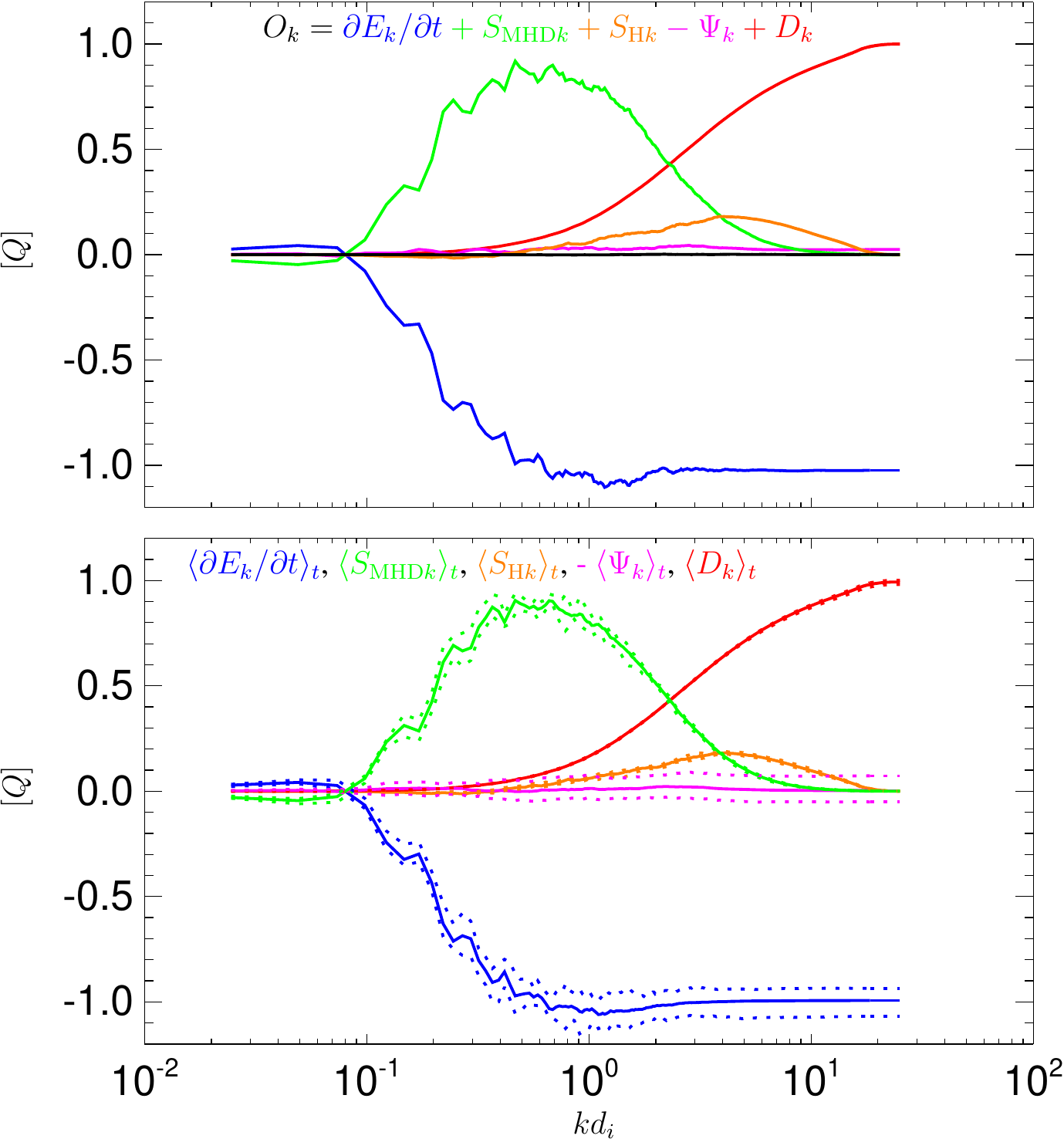}
}
\caption{Spectral transfer analysis:
(top) The validity test of the compressible
ST $O_k$, Eq.~(\ref{ok}),  (black line)  as a function of $k$ at $t=255\Omega_i^{-1}$
along with
the different contributing terms:
(blue) the decay $\partial E_{k}/\partial t$,
(green) the MHD cascade $S_{\text{MHD}k}$,
(orange) the Hall cascade $S_{\text{H}k}$,
(red) the dissipation $D_k$ , and
(magenta) the pressure-dilatation term $-\Psi_k$.
(bottom)  Time-averaged contributing terms (solid lines)
with their minimum and maximum values (dotted lines) for
(blue) the decay $\langle\partial E_{k}/\partial t\rangle_t$,
(green) the MHD cascade $\langle S_{\text{MHD}k}\rangle_t$,
(orange) the Hall cascade $\langle S_{\text{H}k}\rangle_t$,
(red) the dissipation $\langle  D_k\rangle_t$ , and
(magenta) the pressure-dilatation term $-\langle\Psi_k\rangle_t$.
All the quantities are given in units of the total heating rate
$Q$.
\label{avst}}
\end{figure}

To check the spectral transfer of energy given by
Eq.~(\ref{sptrdyn}), we define the validity test (similarly to Eq.~(\ref{oo})) as 
\begin{align}
O_k = \frac{\partial E_{k}}{\partial t} + S_{\text{MHD}k}+S_{\text{H}k} - \Psi_k + D_k
\label{ok}
\end{align}
Fig.~\ref{avst} (top panel) shows the validity test and
the contributing terms as functions of $k$ at $t=255\Omega_i^{-1}$.
Eq.~(\ref{sptrdyn}) is well satisfied, $|O_k|/Q < 0.3~\%$.   
At large scales, small positive values of $\partial E_{k}/\partial t$
compensated by  $S_{\text{MHD}k}$ indicate an inverse cascade/energy transfer
from small to large scales. At medium scales, the MHD cascade term
dominates, reaching the maximum value about $0.85 Q$;
the MHD cascade term is compensated mostly by $\partial E_{k}/\partial t$.
At small scales, the Hall cascade sets in but, at the same
time, the dissipation gets important and a weak pressure-dilatation
effect appears.       
For large $k$ one recovers the energy conservation 
\begin{align}
 \frac{\partial ( E_{\text{kin}}+E_{\text{mag}})}{\partial t} = 
  \langle p\theta \rangle - Q.
\end{align}

As in the KHM case, it is interesting to look at the properties
of the ST equation averaged over one period of the pressure-dilatation
oscillation. Fig.~\ref{avst} (bottom panel) displays 
the terms contributing to the ST equation (Eq.~\ref{sptrdyn})
averaged over the time interval $240 \le t \Omega_i \le  264 $
along with their minimum and maximum values. 
The pressure-dilatation term $\Psi_k$ exhibits a variation
of few percents of $Q$ and, on average, it is negligible.
The decay $\partial E_{k}/\partial t$ and
 the MHD cascade $S_{\text{MHD}k}$ have also a similar
variability whereas the Hall cascade $S_{\text{H}k}$
and the dissipation $D_k$ tend to be about constant
during the oscillation period. 

\begin{figure}
\centerline{\includegraphics[width=8cm]{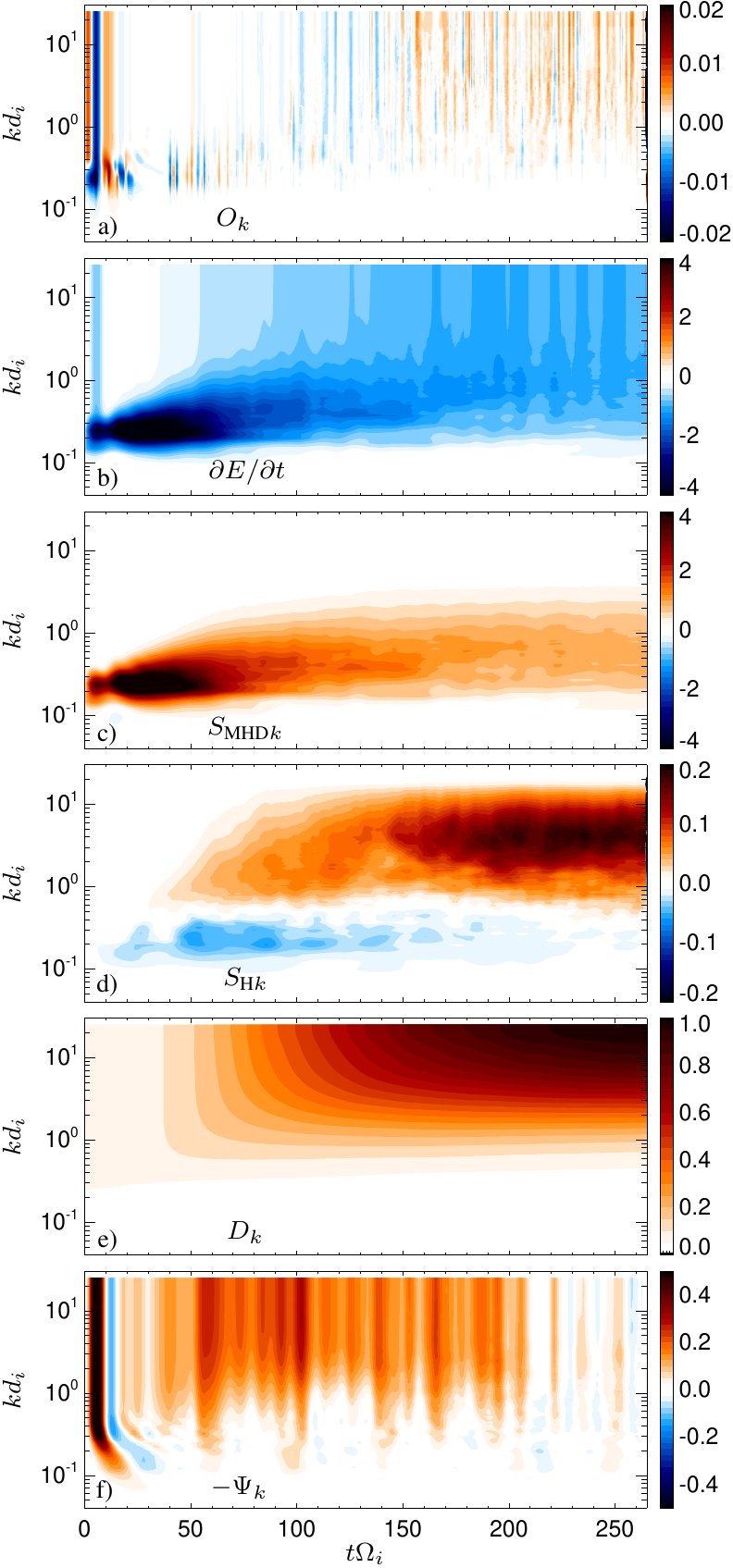}}
\caption{Evolution of the compressible ST law:
color scale plots of
(a) $O_k$, (b) $\partial E_k/\partial t/4$, (c)
$S_{\text{MHD}k}$, (d) $S_{\text{H}k}$, (e) $D_k$, and
(f) $-\Psi_k$ as functions of time and scale $l$.
Dotted lines denote the injection scale $k d_i =0.2$.
All the terms are normalized with respect to the total heating rate
$Q$.
\label{stevol}
}
\end{figure}

The ST equation is valid during the whole simulation 
owing to the peridic boundary condisions as in the case of
the KHM equation.
Figure~\ref{stevol} displays the evolution of the ST
equation in the simulation in a format similar to that
of Fig.~\ref{yevol}.
Figure~\ref{stevol} shows that the ST and KHM 
equations give similar results.
Note, moreover, that a good
quantitative agreement between the two approaches appears at later times
when turbulence is well developed.

\subsection{Comparison between KHM and ST approaches}
\label{compKHMST}

The KHM and ST equations give similar results.
They are empirically related through the inverse 
proportionality between the wave vector $k$ and the spatial separation length $l$
as $ k l  \simeq \sqrt{2}$.
\begin{figure}[htb]
\centerline{
\includegraphics[width=8cm]{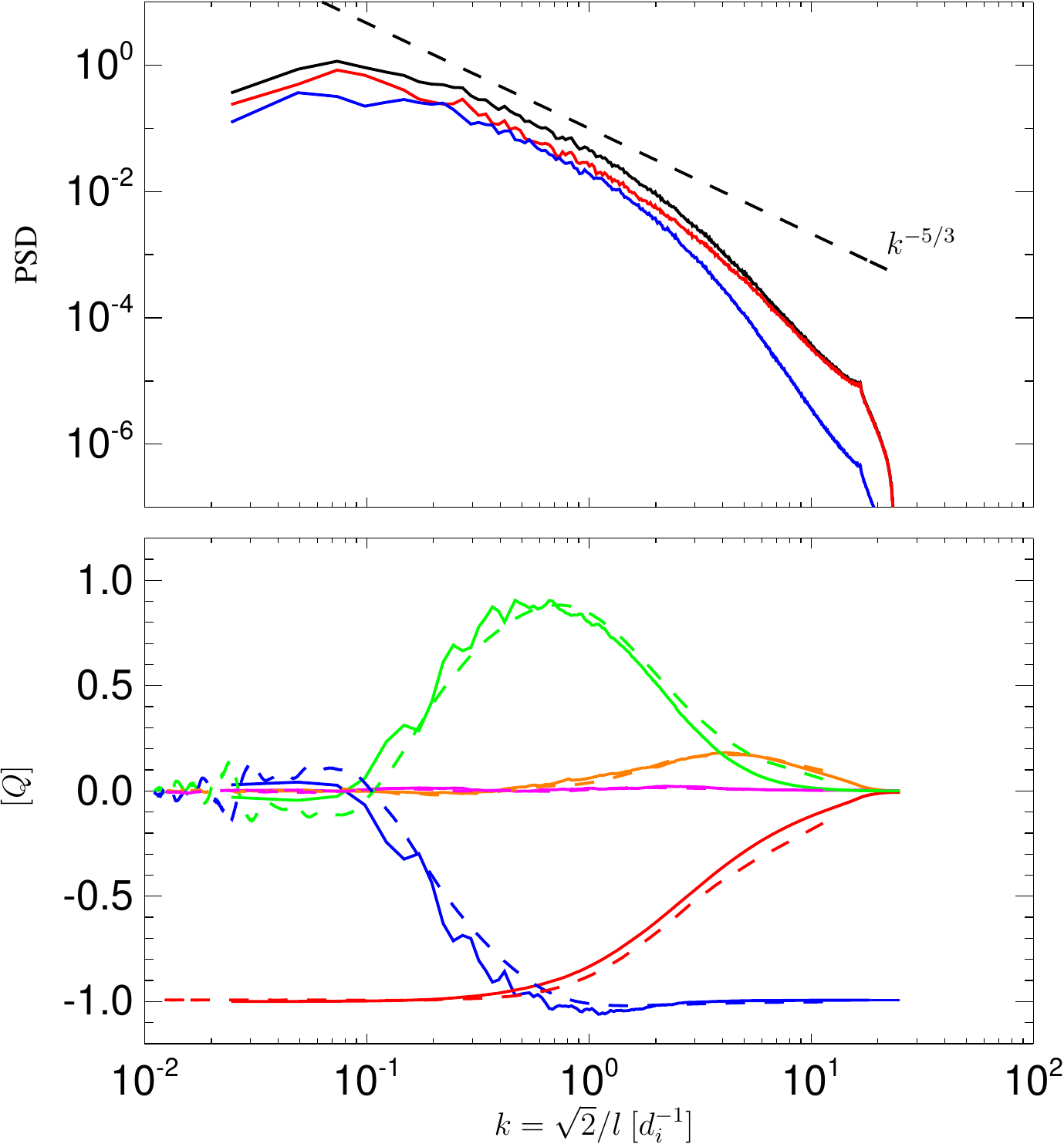}
}
\caption{(top) The power spectra
of the velocity (blue), magnetic field (red),
and the total power spectrum as functions of $k$
at $t= 255  \Omega_i^{-1}$.
(bottom) Solid lines show
the contributing terms of the ST equation 
(averaged over $240 \le t \Omega_i \le  264 $,
see Fig.~\ref{avst}, bottom) as
function of $k$:
(blue) the decay $\langle\partial E_{k}/\partial t\rangle_t$,
(green) the MHD cascade $\langle S_{\text{MHD}k}\rangle_t$,
(orange) the Hall cascade $\langle S_{\text{H}k}\rangle_t$,
(red) the dissipation $\langle  D_k\rangle_t - Q$ , and
(magenta) the pressure-dilatation term $-\langle\Psi_k\rangle_t$.
Dashed lines show the corresponding time-averaged results
of the KHM equation (see Fig.~\ref{yaghall}, bottom) as
function of $k$ through the relation $l=\sqrt{2}/k$:
(blue) the decay $-\langle {\partial \mathcal{S}}/{\partial t}\rangle_t/4-Q$,
(green) the MHD cascade $\langle \mathcal{K}_\text{MHD}\rangle_t$,
(orange) the Hall cascade  $\langle\mathcal{K}_\text{H}\rangle_t$,
(red) the dissipation $-\langle\mathcal{D}\rangle_t$, and
(magenta) the pressure-dilatation $\langle\varPsi\rangle_t$.
All the quantities are given in units of the total
dissipation rate $Q$. Note that some terms are shifted by $-Q$
with respect to Figs.~\ref{yaghall} and~\ref{avst}.
\label{styag}}
\end{figure}
The top panel of Fig.~\ref{styag} shows 
the power spectra
of the velocity (blue), magnetic field (red),
and the total power spectrum as functions of $k$
(at $t= 255  \Omega_i^{-1}$) as a reference.
The dashed line denotes a $k^{-5/3}$ power law for comparison.

The bottom panel Fig.~\ref{styag}
displays a direct comparison between
the ST and KHM contributing terms 
as functions of $k$ (through the inverse
proportionality $ k l = \sqrt{2}$).
Fig.~\ref{styag} shows that the
MHD and Hall cascade rates in both the approaches are
comparable
$\langle S_{\text{MHD}k}\rangle_t \simeq \langle \mathcal{K}_\text{MHD}\rangle_t$,
$\langle S_{\text{H}k}\rangle_t \simeq \langle\mathcal{K}_\text{H}\rangle_t$.
Note that the MHD cascade term dominates at scales where the total
power spectrum is close to a $k^{-5/3}$ power law.
The average pressure-dilation effect is negligible.
Similarly to the hydrodynamic case \citep{hellal21a},
the ST and KHM approaches are complementary:
\begin{align}
\langle\partial E_{k}/\partial t\rangle_t+\langle {\partial \mathcal{S}}/{\partial t}\rangle_t/4 &\simeq -Q \\
\langle  D_k\rangle_t + \langle\mathcal{D}\rangle_t \simeq Q
\end{align}
$\partial E_{k}/{\partial t}$ represents
the rate of change of the kinetic energy at scales
with wave-vector magnitudes smaller or equal to $k$ 
whereas  ${\partial \mathcal{S}}/{\partial t}/4$ represents
approximatively the rate 
for wave-vector magnitudes larger than $k$.
Similarly, $D_{k}$ is the dissipation rate
on the scales $\le k$ whereas $\mathcal{D}$
represents the dissipation rate
on the scales $> k$.

\section{Reduced, one-dimensional analyses}
\label{1Dred}

In the present simulation, due to the reduced
dimensionality, we cannot address the question 
of the spectral anisotropy \citep{verdal15}. We can, however, 
test what estimates of the cascade rate 
we can get from 1D cuts representing observed 1D time series.
We take 1D cuts (both in $x$ and $y$ direction)
to calculate the ST predictions
 $S_\text{MHD}^{(\text{1D})}$ and 
$S_\text{H}^{(\text{1D})}$ 
and the KHM predictions 
 $\mathcal{K}_\text{MHD}^{(\text{1D})}$
and $\mathcal{K}_\text{H}^{(\text{1D})}$.

\begin{figure}[htb]
\centerline{
\includegraphics[width=8cm]{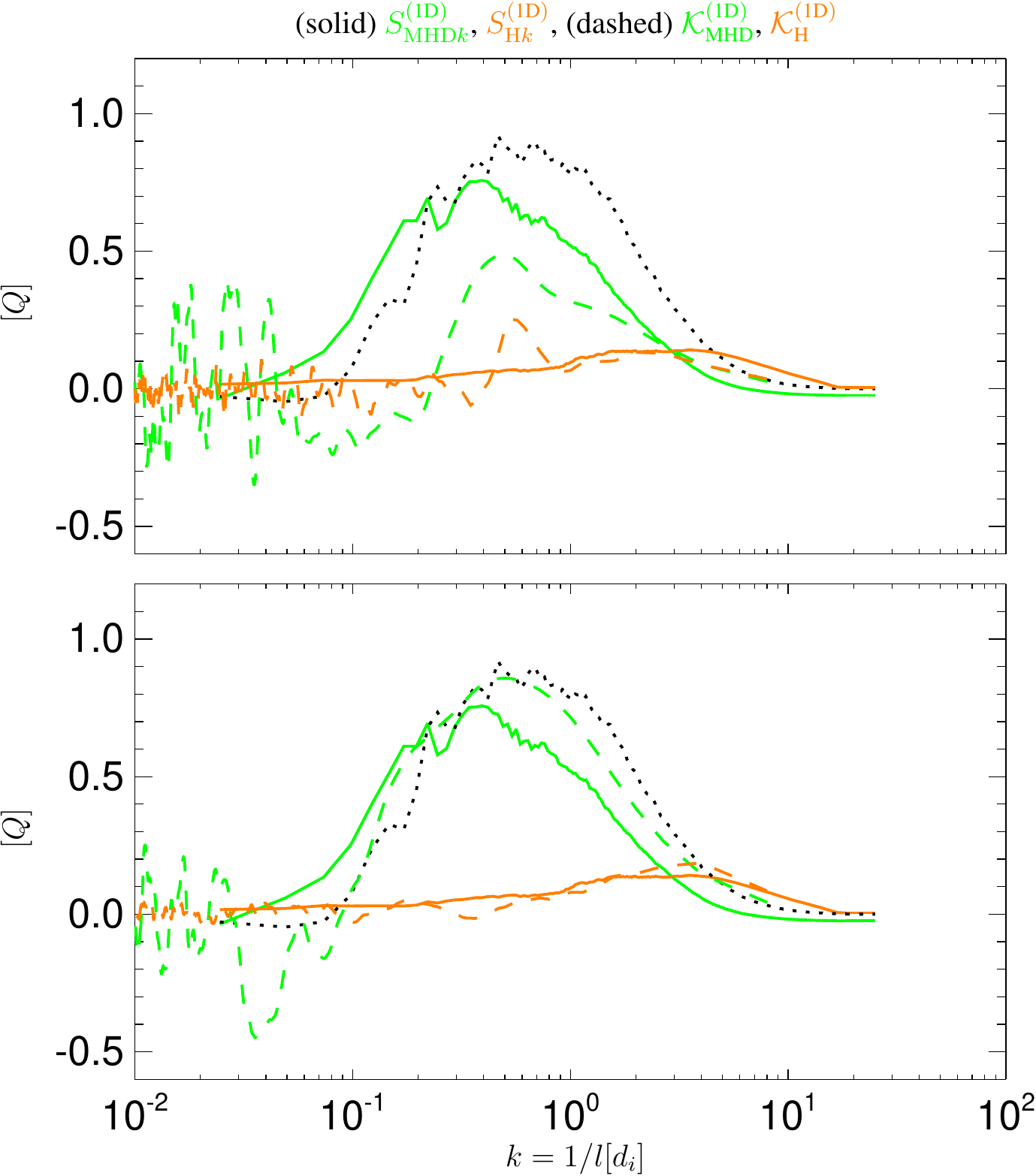}}
\caption{Reduced 1-D analyses of (green) MHD and (orange) Hall cascade rates
(see the text for details):
Solid lines show 
the ST cascade rates $S_\text{MHD}^{(\text{1D})}$
and $S_\text{H}^{(\text{1D})}$ as functions of $k$, estimated from 
1D cuts using Eq.~(\ref{sptrdyn}). Dashed lines displays the
cascade rates $\mathcal{K}_\text{MHD}^{(\text{1D})}$
and $\mathcal{K}_\text{H}^{(\text{1D})}$ as functions of $k=1/l$ 
estimated
from 1D cuts using the standard KHM equation, Eq.~(\ref{CYaglom}), 
in the top panel, whereas in the bottom panel they 
denote the cascade rates estimated
using the alternative  KHM equation,  Eq.~(\ref{CYaglom2}).
Dotted lines displays the MHD cascade rate $S_\text{MHD}$
from the full analysis
(see Fig.~\ref{avst}, top panel) for comparison.  
All the quantities are given in units of the total heating rate
$Q$. 
\label{d1yagst}}
\end{figure}

For the reduced 1D ST equation we take 1D Fourier transform
in Eq.~(\ref{sptrdyn}) and we retain all the spatial derivatives, and,
similarly, we use all the spatial derivatives (in the real space) 
for the alternative KHM equation.
On the other hand,
for the standard KHM equation, Eq.~(\ref{CYaglom}), only the derivative
along the 1D direction in the separation space $l$ is used to estimate
$\boldsymbol{\nabla}\cdot ( \boldsymbol{\mathcal{Y}}+ \boldsymbol{\mathcal{H}})$.
Results from these estimates are shown in 
Fig.~\ref{d1yagst}.
This figure shows the MHD and Hall cascade rates
obtained from the reduced 1D analyses 
as functions of $k$ for the ST approach
 and $k=1/l$ for the KHM approaches.
The calculation is done for the time $t=255 \Omega_i^{-1}$,
see top panels of Figs.~\ref{avst} and~\ref{yaghall}.
The reduced ST equation gives the cascade rates
relatively similar to those obtained from
the whole simulation domain.
On the other hand, the reduced 1D results
based on the standard KHM equation, Eq.~(\ref{CYaglom}),
are quite different from the full results:
the maximum value of the MHD cascade rate is notable about half
the expected value. This can be partly amended by assuming isotropy, 
$\boldsymbol{\nabla}\cdot \boldsymbol{\mathcal{Y}} \sim 2 \partial \mathcal{Y}_l/\partial l$.
However, even after this correction 
the agreement between the 1D ST and standard KHM method is poor.
Interestingly, the reduced 1D results
based on the alternative KHM equation, Eq.~(\ref{CYaglom2}),
are in a good agreement with the full results, as well
as with the results of the reduced 1D ST equation.
This likely related to the fact that for 
the alternative, divergence-less formulation we
use higher-dimensionality derivatives that
correspond to the divergence in the standard KHM equation. 
Finally, we note that the agreement between the 1D ST and KHM results 
appears for $k l = \alpha$, where $\alpha\sim 1$.

\section{Discussion}
\label{discuss}

In this paper we derived two new forms of the KHM equation
for compressible Hall MHD turbulence. We tested
these equations, along with an isotropic ST equation,
on the results of a 2D Hall MHD
simulation of weakly compressible turbulence with
a moderate Reynolds number.
The KHM and ST equations are well satisfied in the
simulation cross-validating these equations and
simulation results.
The two KHM equations give the same results
and they are equivalent and complementary to the ST equation
via the inverse proportionality $k l = \alpha$ with $\alpha \simeq \sqrt{2}$. 
Note that a similar correspondence
is observed in three-dimensional (3D) HD turbulence \citep{hellal21a}
for $\alpha \simeq \sqrt{3}$ and preliminary results
of isotropic (unmagnetized) 3D Hall MHD turbulence
suggest the same relationship. In the reduced, 1D analyses 
we observe $\alpha \simeq 1$. This indicates that the
relationship between the ST and KHM equation $k l = \alpha$  depends on
the dimensionality $d$ of the analyzed space, $\alpha\simeq \sqrt{d}$.
This simple relationship is useful to interpret the KMH
results in the context of spectral analyses.
Equations analogous to the ST and KHM ones can be obtained using
the low-pass filtering/coarse graining of the energy conservation laws
\citep{eyal09,alui11,alui13,yangal16,yangal17b,campal18}.

The KHM and ST equations are valid during the whole simulation,
owing to the periodic boundary conditions and they can be used
to analyze the onset of turbulence. The agreement
between the two approaches is at least qualitative
during the whole simulation but only at later
times, when turbulence is well developed, 
there is a good quantitative agreement; this
indicates that the KHM and ST approaches become
equivalent when turbulence is well developed and the
energy transfer is expected be local so that one can talk about
the energy cascade. The simulation
results suggest that the onset of the Hall cascade is
related to formation of reconnecting current sheets \citep{papial19}
and that, at least during the onset, the Hall term leads to an energy transfer 
from small to large scales
 at a range of scales above $d_i$ \cite[cf.,][]{franal17};
at this stage, this transfer is likely not local
and one cannot speak about a energy cascade.
The locality can be tested using
a (spectral) shell-to-shell transfer analysis \citep{alexal05}.
For the weakly compressible simulation, the incompressible
approximation is not correct to a large extent due to the neglected pressure-dilatation
effect. This error is important during the turbulence onset.  
When turbulence is well developed, the time-averaged pressure-dilatation
becomes negligible, and the compressible and incompressible
KHM predictions for MHD and Hall cascade rates are close to each other.

It is interesting to note that the standard compressible KHM
equation, Eq.~(\ref{CYaglom}), does not depend (except for
the correction term $\mathcal{C}$)
on the background magnetic field $\boldsymbol{B}_0=\langle \boldsymbol{B}\rangle$
similarly to the incompressible case \citep{oughal13}. On the other hand,
there is in principle a contribution from the mean fluid
velocity $\boldsymbol{u}_0=\langle \boldsymbol{u}\rangle$ \citep{hadial17}.
In the incompressible, constant-density limit one
recovers the incompressible results \citep{hellal18,ferral19,baga17}. 

There is a couple of differences between our standard KHM equation (Eq.~(\ref{CYaglom})) and
that of \cite{andral18}:
We describe the kinetic energy by the structure
function $\langle |\delta \boldsymbol{w}|^2 \rangle$ that guarantee positive
values in contrast with $\langle \delta(\rho\boldsymbol{u})\cdot \delta \boldsymbol{u} \rangle$.
Similarly, we represent the magnetic energy contribution  by
the structure function $S_B=\langle |\delta \boldsymbol{B}|^2 \rangle$
instead of $\langle \delta(\rho^{1/2}  \boldsymbol{B}) 
\cdot \delta(\boldsymbol{B}/\rho^{1/2}) \rangle$.
This difference is very likely not substantial. In the case 
of compressible HD turbulence, the corresponding two approaches
give very similar results \citep{hellal21a}.
However, the important difference between our results and those of \cite{andral18} 
 is that we formulate the KHM equations for the kinetic + magnetic energy. 
The simulation results exhibit no net energy exchange between the kinetic + magnetic energy 
and the internal one for later times when turbulence is well developed
and the pressure-dilation effect does not lead to cross-scale energy
transfer \cite[cf.,][]{aluial12}.
The inclusion of the internal energy into the KHM equation
\cite[][and references therein]{andral18} is therefore not needed.
Moreover, the HD results of \cite{hellal21a} indicate that it
is hard to represent the kinetic energy and the internal one
by compatible structure functions. Furthermore, \cite{andral18} 
use the isothermal closure to modify the description of the
pressure-dilatation effect; this closure is not generally
applicable. Finally,  estimates of the energy cascade rate in the
solar wind based on such approaches 
\citep{baneal16,hadial17,andral19} include
the scale redistribution of the internal energy
and cannot be simply related to the plasma heating rates
\cite[cf.,][]{macbal08}.

For our study, we used one 2D Hall MHD simulation of weakly
compressible turbulence for a moderated Reynolds number and
zero cross-helicity.  This was sufficient
for testing and comparing the KHM and ST equations, but
our results need to be extended to more
compressible and/or larger system that extends further
to the Hall range of scales, and/or to systems 
with larger cross (canonical) helicity, etc. 
Our results need to be extended to 3D 
in order to investigate the anisotropy
of the turbulent cascade.  Both the KHM equations
can be naturally used \cite[cf.,][]{verdal15}. However, it is not clear
how to extend the isotropic ST approach to an
anisotropic situation \cite[cf.,][]{verm17};
the low-pass filtering/coarse graining approaches also
usually assume isotropy.

The Hall cascade rate in the simulation is a fraction of
the MHD one because of the dissipation. In
both kinetic simulations and in-situ observations the situation is similar
\citep{bandal20b}. The decrease of
the cascade rate from the MHD to the Hall range
is likely due to some sort of dissipation/particle energization.
The present work assumes a scalar pressure $p$
that is relevant for collision-dominated plasmas.  
In weakly collisional/collisionless systems,
such as in the solar wind, it is
necessary to employ the full pressure tensor $\boldsymbol{\mathrm{P}}$
and to replace the pressure-dilatation coupling, $p\theta$,
by the pressure-strain one, $\boldsymbol{\mathrm{P}}:\boldsymbol{\Sigma}$,
\citep{yangal19,mattal20}. The KHM and ST approaches
can be easily generalized to account for the pressure tensor.
For instance, $\langle \delta p \delta \theta \rangle$ becomes
simply $\langle \delta \boldsymbol{\mathrm{P}}  : \delta \boldsymbol{\Sigma} \rangle$
(see the viscous dissipation term).

Our reduced 1D analysis suggests that the ST equation and the
alternative KHM equation give better estimations
of the cascade rate from in situ observed time series compared to the
usually used standard KHM. However, this is likely owing to the usage
of multi-dimensional spatial derivatives
that may possibly be only estimated using
multi-point data of Cluster or MMS missions.
Furthermore, the ST equation is formulated
for an isotropic situation. Nevertheless,
we believe that the ST and alternative
KHM approaches are worth pursuing as
they provide other methods for measuring
the energy cascade rates from in situ
observations.

\acknowledgments
P.H. acknowledges grant 18-08861S of the Czech Science Foundation.
This research was conducted with high performance computing (HPC) resources provided
by the Cineca ISCRA initiative (grant HP10C2EARF).
L.F. is supported by the UK Science and Technology Facilities Council (STFC) grant ST/T00018X/1.

\appendix

\section{Derivation of the standard KHM equation}

We start by reformulating the compressible Hall MHD equation
in terms of $\boldsymbol{w}= \rho^{1/2} \boldsymbol{u}$.
\begin{align}
\frac{\partial\boldsymbol{w}}{\partial t}&+(\boldsymbol{u}\cdot\boldsymbol{\nabla})\boldsymbol{w}
+\frac{\boldsymbol{w}\theta }{2}= \frac{\boldsymbol{f}}{\sqrt{\rho}}
\\
\frac{\partial\boldsymbol{B}}{\partial t}&+(\boldsymbol{u}\cdot\boldsymbol{\nabla})\boldsymbol{B}=(\boldsymbol{B}\cdot\boldsymbol{\nabla})\boldsymbol{u}-\boldsymbol{B} \theta +\eta \Delta \boldsymbol{B} +\boldsymbol{h}
\end{align}
where we denoted 
\begin{align}
\boldsymbol{f}&=(\boldsymbol{\nabla}\times\boldsymbol{B})\times\boldsymbol{B}-\boldsymbol{\nabla}p+\boldsymbol{\nabla}\cdot\boldsymbol{\tau} \\
\boldsymbol{h}&=-(\boldsymbol{B}\cdot\boldsymbol{\nabla})\boldsymbol{j}+(\boldsymbol{j}\cdot\boldsymbol{\nabla})\boldsymbol{B} +\boldsymbol{B}\left(\boldsymbol{\nabla}\cdot\boldsymbol{j}\right) .
\end{align}
We take the Hall MHD equations at two different points,
 $\boldsymbol{x}^\prime=\boldsymbol{x}+\boldsymbol{l}$ 
and $\boldsymbol{x}$ and subtract them to get
\begin{align}
\frac{\partial\delta\boldsymbol{w}}{\partial t}&+(\delta\boldsymbol{u}\cdot\boldsymbol{\nabla}^{\prime})\delta\boldsymbol{w}+\left[\boldsymbol{u}\cdot\left(\boldsymbol{\nabla}+\boldsymbol{\nabla}^{\prime}\right)\right]\delta\boldsymbol{w} \nonumber \\ &+\frac{1}{2}(\boldsymbol{w}^\prime\theta^\prime - \boldsymbol{w}\theta)
= \delta\left(\frac{\boldsymbol{f}}{\sqrt{\rho}}\right)
\end{align}
where we denote $a=a(\boldsymbol{x})$, $a^\prime=a(\boldsymbol{x}^\prime)$, and
$\delta a = a^\prime - a$ for any variable $a$; analogically, 
 $\boldsymbol{\nabla}=\boldsymbol{\nabla}_{\boldsymbol{x}}$, 
 $\boldsymbol{\nabla}^\prime=\boldsymbol{\nabla}_{\boldsymbol{x}^\prime}$
and $\Delta=\Delta_{\boldsymbol{x}}$, $\Delta^\prime=\Delta_{\boldsymbol{x}^\prime}$,
$\Delta$ being the Laplace operator. 
Here and henceforth this relationship is repeatedly applied \citep{carbal09}
\begin{align}
\delta \left[ (\boldsymbol{a}\cdot\boldsymbol{\nabla}) \boldsymbol{b}  \right] &=
 (\boldsymbol{a}^{\prime}\cdot\boldsymbol{\nabla}^{\prime}) \boldsymbol{b}^{\prime} - (\boldsymbol{a}\cdot\boldsymbol{\nabla}) \boldsymbol{b} \nonumber \\
&= (\boldsymbol{a}^{\prime}\cdot\boldsymbol{\nabla}^{\prime}) \delta \boldsymbol{b} + (\boldsymbol{a}\cdot\boldsymbol{\nabla}) \delta \boldsymbol{b} \nonumber \\
&= (\delta\boldsymbol{a}\cdot\boldsymbol{\nabla}^{\prime}) \delta \boldsymbol{b} + 
[\boldsymbol{a}\cdot(\boldsymbol{\nabla} +\boldsymbol{\nabla}^{\prime} )] \delta \boldsymbol{b},
\end{align}
and $\boldsymbol{x}$ and $\boldsymbol{x}^\prime$ are assumed to be independent, 
i.e., $\boldsymbol{\nabla}^{\prime}\boldsymbol{a}=\boldsymbol{\nabla}\boldsymbol{a}^\prime=0$ 
for any $\boldsymbol{a}$.
\begin{align}
\frac{\partial|\delta\boldsymbol{w}|^{2}}{\partial t}&+
\boldsymbol{\nabla}^{\prime}\cdot\left(\delta\boldsymbol{u}|\delta\boldsymbol{w}|^{2}\right)
+\left(\boldsymbol{\nabla}
+\boldsymbol{\nabla}^{\prime}\right)\cdot\left(\boldsymbol{u}|\delta\boldsymbol{w}|^{2}\right) 
\\
&+\delta\boldsymbol{w}\cdot\left(\theta^{\prime}\boldsymbol{w}-\theta\boldsymbol{w}^{\prime}\right) = 2 \delta\boldsymbol{w} \cdot \delta\left(\frac{\boldsymbol{f}}{\sqrt{\rho}}\right)
\end{align}
Taking a spatial average 
\begin{align}
\frac{\partial \mathcal{S}_{w}}{\partial t}+\boldsymbol{\nabla}_{\boldsymbol{l}}\cdot\left\langle \delta\boldsymbol{u}|\delta\boldsymbol{w}|^{2}\right\rangle +\mathcal{R}=2\left\langle \delta\boldsymbol{w}\cdot\delta
\left(\frac{\boldsymbol{f}}{\sqrt{\rho}}\right)\right\rangle 
\end{align}
where $\mathcal{S}_{w}=\langle |\delta\boldsymbol{w}|^{2}\rangle$ and 
$\mathcal{R}=\langle \delta\boldsymbol{w}\cdot\left(\theta^{\prime}\boldsymbol{w}-\theta\boldsymbol{w}^{\prime}\right) \rangle$.
Here we assume that the system is homogeneous, i.e., 
\begin{align}
\langle \left(\boldsymbol{\nabla}
+\boldsymbol{\nabla}^{\prime}\right)\cdot \boldsymbol{a}\rangle =0 
\end{align}
for any quantity $\boldsymbol{a}$ \citep{fris95}.

Similarly for the magnetic field we have
\begin{align}
\frac{\partial|\delta\boldsymbol{B}|^{2}}{\partial t} 
&+\boldsymbol{\nabla}^{\prime}\cdot\left(\delta\boldsymbol{u}|\delta\boldsymbol{B}|^{2}\right)
+\left(\boldsymbol{\nabla}+\boldsymbol{\nabla}^{\prime}\right)\cdot\left(\boldsymbol{u}|\delta\boldsymbol{B}|^{2}\right) \nonumber \\ &=
2\delta\boldsymbol{B}\cdot(\delta\boldsymbol{B}\cdot\boldsymbol{\nabla}^{\prime})\delta\boldsymbol{u}
+ \left(B^{\prime2}-B^{2}\right)\left(\theta-\theta^{\prime}\right)
 \nonumber \\
&+2\delta\boldsymbol{B}\cdot\left[\boldsymbol{B}\cdot\left(\boldsymbol{\nabla}+\boldsymbol{\nabla}^{\prime}\right)\right])\delta\boldsymbol{u}
\nonumber \\
&+ 2\eta\delta\boldsymbol{B}\cdot\left(\Delta^{\prime}\boldsymbol{B}^{\prime}-\Delta\boldsymbol{B}\right)
+2 \delta\boldsymbol{B}\cdot\delta \boldsymbol{h}
\end{align}
Taking the spatial average
\begin{align}
\frac{\partial \mathcal{S}_{B}}{\partial t}&
+\boldsymbol{\nabla}_{\boldsymbol{l}}\cdot\left\langle \delta\boldsymbol{u}|\delta\boldsymbol{B}|^{2}\right\rangle 
-2\left\langle \delta\boldsymbol{B}\cdot(\delta\boldsymbol{B}\cdot\boldsymbol{\nabla}^{\prime})\delta\boldsymbol{u}\right\rangle \nonumber \\
&=-\left\langle \delta\left(B^{2}\right)\delta\theta \right\rangle 
+2\left\langle\delta\boldsymbol{B}\cdot\left[\boldsymbol{B}\cdot\left(\boldsymbol{\nabla}+\boldsymbol{\nabla}^{\prime}\right)\right])\delta\boldsymbol{u} \right\rangle \nonumber \\ 
& -4 Q_\eta +2\eta\Delta_{\boldsymbol{l}} S_{B}+2\left\langle \delta\boldsymbol{B}\cdot\delta\boldsymbol{h}\right\rangle 
\end{align}
where $S_{B}=\langle|\delta\boldsymbol{B}|^{2}\rangle$, $Q_\eta =\eta \langle\boldsymbol{\nabla} \boldsymbol{B}:\boldsymbol{\nabla} \boldsymbol{B} \rangle = \eta \langle J^2 \rangle.$
In order to derive an equation for the structure function
$\mathcal{S}=\mathcal{S}_w+\mathcal{S}_B$
and to obtain a simple equation we introduce a correction term
$\mathcal{C}_\text{MHD}=2 
\langle \delta\boldsymbol{w} \cdot \delta\left(\rho^{-1/2}\boldsymbol{f}\right)\rangle
-2 \langle \delta\boldsymbol{u} \cdot  \delta \boldsymbol{f}  \rangle.$
The last term can be expressed as
\begin{align}
2\left\langle \delta\boldsymbol{u}\cdot\delta\boldsymbol{f}\right\rangle &= 
2\left\langle \delta\boldsymbol{u}\cdot\left(\boldsymbol{B}\cdot\left(\boldsymbol{\nabla}+\boldsymbol{\nabla}^{\prime}\right)\right)\delta\boldsymbol{B}\right\rangle 
\nonumber \\
&+ 2\left\langle \delta\boldsymbol{u}\cdot\left(\delta\boldsymbol{B}\cdot\boldsymbol{\nabla}^{\prime}\right)\delta\boldsymbol{B}\right\rangle 
-\left\langle \delta\boldsymbol{u}\cdot\delta\left(\boldsymbol{\nabla}B^{2}\right)\right\rangle 
\nonumber \\
&+2\left\langle \delta\theta\delta p\right\rangle -2\left\langle \delta\boldsymbol{\Sigma}:\delta\boldsymbol{\tau}\right\rangle.
\end{align}
Combining the previous results we get for $\mathcal{S}$
\begin{align}
\frac{\partial \mathcal{S}}{\partial t}+\boldsymbol{\nabla}_{\boldsymbol{l}}\cdot\boldsymbol{\mathcal{Y}}
+\mathcal{R}&=\mathcal{C}_\text{MHD}
+2\left\langle \delta\theta\delta p\right\rangle -2\left\langle \delta\boldsymbol{\Sigma}:\delta\boldsymbol{\tau}\right\rangle  \nonumber \\
& -4 Q_\eta +2\eta\Delta_{\boldsymbol{l}} \mathcal{S}_{B}+2 \langle \delta\boldsymbol{B}\cdot\delta \boldsymbol{h} \rangle
\end{align}
where $\boldsymbol{\mathcal{Y}}=\left\langle \delta\boldsymbol{u}
\left(
|\delta\boldsymbol{w}|^{2}+|\delta\boldsymbol{B}|^{2}
\right) - 2\delta \boldsymbol{B} \left( \delta \boldsymbol{B}\cdot \delta \boldsymbol{u} \right)
\right\rangle
$

To finish the calculation we need to evaluated the Hall contribution $2 \langle \delta\boldsymbol{B}\cdot\delta \boldsymbol{h} \rangle$. First, for $\delta \boldsymbol{h}$ is easy to get
\begin{align}
\delta \boldsymbol{h}&= 
-(\delta\boldsymbol{B}\cdot\boldsymbol{\nabla}^{\prime})\delta\boldsymbol{j} 
-(\boldsymbol{B}\cdot\left(\boldsymbol{\nabla}+\boldsymbol{\nabla}^{\prime}\right))\delta\boldsymbol{j} \nonumber \\
&+(\delta\boldsymbol{j}\cdot\boldsymbol{\nabla}^{\prime})\delta\boldsymbol{B}+(\boldsymbol{j}\cdot\left(\boldsymbol{\nabla}+\boldsymbol{\nabla}^{\prime}\right))\delta\boldsymbol{B} \nonumber \\
&+\delta\boldsymbol{B}\left(\boldsymbol{\nabla}^{\prime}\cdot\delta\boldsymbol{j}\right)+\boldsymbol{B}\left(\left[\boldsymbol{\nabla}+\boldsymbol{\nabla}^{\prime}\right]\cdot\delta\boldsymbol{j}\right)
\end{align}
The term 
$2 \langle \delta\boldsymbol{B}\cdot\delta \boldsymbol{h} \rangle$
may be then given in the form \citep{hellal18,ferral19}
\begin{align}
2\left\langle \delta\boldsymbol{B}\cdot\delta\boldsymbol{h}\right\rangle =
-2\boldsymbol{\nabla}_{\boldsymbol{l}}\cdot\boldsymbol{\mathcal{H}}+2 \mathcal{A}
\end{align}
where 
\begin{align}
\boldsymbol{\mathcal{H}}&=\left\langle \delta\boldsymbol{B}\left(\delta\boldsymbol{B}\cdot\delta\boldsymbol{j}\right)-\delta\boldsymbol{j}\left|\delta\boldsymbol{B}\right|^{2}/2\right\rangle \nonumber \\
&=\left\langle \delta\boldsymbol{B}\times\left(\delta\boldsymbol{B}\times\delta\boldsymbol{j}\right)+\delta\boldsymbol{j}\left|\delta\boldsymbol{B}\right|^{2}/2\right\rangle 
\end{align}
and
\begin{align}
 \mathcal{A}=\left\langle \boldsymbol{j}\cdot\left(\boldsymbol{B}^{\prime}\times\boldsymbol{J}^{\prime}\right)+\boldsymbol{j}^{\prime}\cdot\left(\boldsymbol{B}\times\boldsymbol{J}\right)\right\rangle.
\end{align}
The term $\mathcal{A}$ can be expressed as
\begin{align}
\mathcal{A}&=\mathcal{C}_\text{H}+\left\langle \boldsymbol{J}\cdot\left(\boldsymbol{B}^{\prime}\times\boldsymbol{j}^{\prime}\right)+\boldsymbol{J}^{\prime}\cdot\left(\boldsymbol{B}\times\boldsymbol{j}\right)\right\rangle \\
&=\mathcal{C}_\text{H}+\boldsymbol{\nabla}^{\prime}\cdot\left[\left\langle \delta\boldsymbol{B}\times\left(\boldsymbol{B}^{\prime}\times\boldsymbol{j}^{\prime}\right)+\delta\boldsymbol{B}\times\left(\boldsymbol{B}\times\boldsymbol{j}\right)\right\rangle \right] \nonumber
\label{A1}
\end{align}
where
\begin{align}
\mathcal{C}_\text{H}&=\left\langle \boldsymbol{j}\cdot\left(\boldsymbol{B}^{\prime}\times\boldsymbol{J}^{\prime}\right)+\boldsymbol{j}^{\prime}\cdot\left(\boldsymbol{B}\times\boldsymbol{J}\right)\right\rangle  \nonumber \\
 &-\left\langle \boldsymbol{J}\cdot\left(\boldsymbol{B}^{\prime}\times\boldsymbol{j}^{\prime}\right)+\boldsymbol{J}^{\prime}\cdot\left(\boldsymbol{B}\times\boldsymbol{j}\right)\right\rangle.
\end{align}
It can also be transformed into this form
\begin{align}
\mathcal{A}&=-\left\langle \boldsymbol{J}\cdot\left(\boldsymbol{B}^{\prime}\times\boldsymbol{j}^{\prime}\right)+\boldsymbol{J}^{\prime}\cdot\left(\boldsymbol{B}\times\boldsymbol{j}\right)\right\rangle \nonumber \\
&=-\boldsymbol{\nabla}^{\prime}\cdot\left\langle \delta\boldsymbol{B}\times\left(\boldsymbol{B}^{\prime}\times\boldsymbol{j}\right)\right\rangle -\boldsymbol{\nabla}^{\prime}\cdot\left\langle \delta\boldsymbol{B}\times\left(\boldsymbol{B}\times\boldsymbol{j}^{\prime}\right)\right\rangle \nonumber \\
&-\left\langle \left(\boldsymbol{j}^{\prime}\cdot\boldsymbol{\nabla}\right)|\delta\boldsymbol{B}|^{2}/2\right\rangle -\left\langle \left(\boldsymbol{j}\cdot\boldsymbol{\nabla}^{\prime}\right)|\delta\boldsymbol{B}|^{2}/2\right\rangle .
\label{A2}
\end{align}
Combining Eq.~(\ref{A1}) and Eq.~(\ref{A2}) gives 
\begin{align}
2 \mathcal{A} = \mathcal{C}_\text{H} + \boldsymbol{\nabla}_{\boldsymbol{l}}\cdot\boldsymbol{\mathcal{H}}
\end{align}
The final version of the KHM equation in compressible Hall MHD gets
this form
\begin{align}
\frac{\partial \mathcal{S}}{\partial t}+
\boldsymbol{\nabla}_{\boldsymbol{l}}\cdot\left(\boldsymbol{\mathcal{Y}}+\boldsymbol{\mathcal{H}}\right) 
 +\mathcal{R}&= \mathcal{C}
+2\left\langle \delta\theta\delta p\right\rangle -2\left\langle \delta\boldsymbol{\Sigma}:\delta\boldsymbol{\tau}\right\rangle  \nonumber \\
& -4 Q_\eta +2\eta\Delta_{\boldsymbol{l}} \mathcal{S}_{B}
\end{align}
with $\mathcal{C}=\mathcal{C}_\text{MHD}+\mathcal{C}_\text{H}$.


\begin{thebibliography}{70}
\expandafter\ifx\csname natexlab\endcsname\relax\def\natexlab#1{#1}\fi

\bibitem[{{Alexakis} {et~al.}(2005){Alexakis}, {Mininni}, \&
  {Pouquet}}]{alexal05}
{Alexakis}, A., {Mininni}, P.~D., \& {Pouquet}, A. 2005, Phys. Rev. E, 72,
  046301

\bibitem[{Aluie(2011)}]{alui11}
Aluie, H. 2011, Phys. Rev. Lett., 106, 174502

\bibitem[{Aluie(2013)}]{alui13}
---. 2013, Physica D, 247, 54

\bibitem[{Aluie {et~al.}(2012)Aluie, Li, \& Li}]{aluial12}
Aluie, H., Li, S., \& Li, H. 2012, ApJL, 751, L29

\bibitem[{Andr{\'e}s {et~al.}(2018)Andr{\'e}s, Galtier, \& Sahraoui}]{andral18}
Andr{\'e}s, N., Galtier, S., \& Sahraoui, F. 2018, Phys. Rev. E, 97, 013204

\bibitem[{Andr{\'e}s {et~al.}(2019)Andr{\'e}s, Sahraoui, Galtier, Hadid,
  Ferrand, \& Huang}]{andral19}
Andr{\'e}s, N., Sahraoui, F., Galtier, S., Hadid, L.~Z., Ferrand, R., \& Huang,
  S.~Y. 2019, Phys. Rev. Lett., 123, 245101

\bibitem[{Bandyopadhyay {et~al.}(2020{\natexlab{a}})Bandyopadhyay, Goldstein,
  Maruca, Matthaeus, Parashar, Ruffolo, Chhiber, Usmanov, Chasapis, Qudsi,
  Bale, Bonnell, {Dudok de Wit}, {Goetz}, Harvey, MacDowall, Malaspina, Pulupa,
  Kasper, Korreck, Case, Stevens, Whittlesey, Larson, Livi, Klein, Velli, \&
  Raouafi}]{bandal20a}
Bandyopadhyay, R., {et~al.} 2020{\natexlab{a}}, ApJS, 246, 48

\bibitem[{Bandyopadhyay {et~al.}(2020{\natexlab{b}})Bandyopadhyay,
  Sorriso-Valvo, Chasapis, Hellinger, Matthaeus, Verdini, Landi, Franci,
  Matteini, Giles, Gershman, Pollock, Russell, Strangeway, Torbert, Moore, \&
  Burch}]{bandal20b}
---. 2020{\natexlab{b}}, Phys. Rev. Lett., 124, 225101

\bibitem[{Banerjee \& Galtier(2017)}]{baga17}
Banerjee, S., \& Galtier, S. 2017, J. Phys. A, 50, 015501

\bibitem[{{Banerjee} {et~al.}(2016){Banerjee}, {Hadid}, {Sahraoui}, \&
  {Galtier}}]{baneal16}
{Banerjee}, S., {Hadid}, L.~Z., {Sahraoui}, F., \& {Galtier}, S. 2016, ApJL,
  829, L27

\bibitem[{Bruno \& Carbone(2013)}]{brca13}
Bruno, R., \& Carbone, V. 2013, LRSP, 10, 2

\bibitem[{Camporeale {et~al.}(2018)Camporeale, Sorriso-Valvo, Califano, \&
  Retin{\`o}}]{campal18}
Camporeale, E., Sorriso-Valvo, L., Califano, F., \& Retin{\`o}, A. 2018, Phys.
  Rev. Lett., 120, 125101

\bibitem[{Carbone {et~al.}(2009)Carbone, Sorriso-Valvo, \& Marino}]{carbal09}
Carbone, V., Sorriso-Valvo, V., \& Marino, R. 2009, Eur. Phys. Lett., 88, 25001

\bibitem[{Chen {et~al.}(2013)Chen, Boldyrev, Xia, \& Perez}]{chenal13b}
Chen, C. H.~K., Boldyrev, S., Xia, Q., \& Perez, J.~C. 2013, Phys. Rev. Lett.,
  110, 225002

\bibitem[{{Chen} {et~al.}(2014){Chen}, {Leung}, {Boldyrev}, {Maruca}, \&
  {Bale}}]{chenal14}
{Chen}, C.~H.~K., {Leung}, L., {Boldyrev}, S., {Maruca}, B.~A., \& {Bale},
  S.~D. 2014, Geophys. Res. Lett., 41, 8081

\bibitem[{Coburn {et~al.}(2015)Coburn, Forman, Smith, Vasquez, \&
  Stawarz}]{cobual15}
Coburn, J.~T., Forman, M.~A., Smith, C.~W., Vasquez, B.~J., \& Stawarz, J.~E.
  2015, Phil. Trans. R. Soc. A, 373, 20140150

\bibitem[{Cranmer {et~al.}(2009)Cranmer, Matthaeus, Breech, \&
  Kasper}]{cranal09}
Cranmer, S.~R., Matthaeus, W.~H., Breech, B.~A., \& Kasper, J.~C. 2009, ApJ,
  702, 1604

\bibitem[{{de K\'arm\'an} \& {Howarth}(1938)}]{kaho38}
{de K\'arm\'an}, T., \& {Howarth}, L. 1938, Proc. Royal Soc. London Series A,
  164, 192

\bibitem[{Eyink \& Aluie(2009)}]{eyal09}
Eyink, G.~L., \& Aluie, H. 2009, Phys. Fluids, 21, 115107

\bibitem[{Ferrand {et~al.}(2019)Ferrand, Galtier, Sahraoui, Meyrand,
  {Andr\'es}, \& Banerjee}]{ferral19}
Ferrand, R., Galtier, S., Sahraoui, F., Meyrand, R., {Andr\'es}, N., \&
  Banerjee, S. 2019, ApJ, 881, 50

\bibitem[{Franci {et~al.}(2016)Franci, Landi, Matteini, Verdini, \&
  Hellinger}]{franal16b}
Franci, L., Landi, S., Matteini, L., Verdini, A., \& Hellinger, P. 2016, ApJ,
  833, 91

\bibitem[{Franci {et~al.}(2017)Franci, Cerri, Califano, Landi, Papini, Verdini,
  Matteini, Jenko, \& Hellinger}]{franal17}
Franci, L., {et~al.} 2017, ApJL, 850, L16

\bibitem[{Franci {et~al.}(2020)Franci, Stawarz, Papini, Hellinger, Nakamura,
  Burgess, Landi, Verdini, Matteini, Ergun, {Le Contel}, \&
  Lindqvist}]{franal20}
---. 2020, ApJ, 898, 175

\bibitem[{Frisch(1995)}]{fris95}
Frisch, U. 1995, Turbulence (Cambridge University Press)

\bibitem[{{Galtier}(2008)}]{galt08}
{Galtier}, S. 2008, Phys. Rev. E, 77, 015302

\bibitem[{Ghosh {et~al.}(1993)Ghosh, Hossain, \& Matthaeus}]{ghosal93}
Ghosh, S., Hossain, M., \& Matthaeus, W.~H. 1993, Comp. Phys. Comm., 74, 18

\bibitem[{{Grete} {et~al.}(2017){Grete}, {O'Shea}, {Beckwith}, {Schmidt}, \&
  {Christlieb}}]{gretal17}
{Grete}, P., {O'Shea}, B.~W., {Beckwith}, K., {Schmidt}, W., \& {Christlieb},
  A. 2017, Phys. Plasmas, 24, 092311

\bibitem[{{Hadid} {et~al.}(2017){Hadid}, {Sahraoui}, \& {Galtier}}]{hadial17}
{Hadid}, L.~Z., {Sahraoui}, F., \& {Galtier}, S. 2017, ApJ, 838, 9

\bibitem[{Hellinger {et~al.}(2011)Hellinger, Matteini, {\v S}tver\'ak,
  {Tr\'avn\'{\i}\v{c}ek}, \& Marsch}]{hellal11}
Hellinger, P., Matteini, L., {\v S}tver\'ak, {\v S}., {Tr\'avn\'{\i}\v{c}ek},
  P.~M., \& Marsch, E. 2011, J. Geophys Res., 116, A09105

\bibitem[{Hellinger {et~al.}(2013)Hellinger, {Tr\'avn\'{\i}\v{c}ek}, {\v
  S}tver\'ak, Matteini, \& Velli}]{hellal13}
Hellinger, P., {Tr\'avn\'{\i}\v{c}ek}, P.~M., {\v S}tver\'ak, {\v S}.,
  Matteini, L., \& Velli, M. 2013, J. Geophys Res., 118, 1351

\bibitem[{Hellinger {et~al.}(2018)Hellinger, Verdini, Landi, Franci, \&
  Matteini}]{hellal18}
Hellinger, P., Verdini, A., Landi, S., Franci, L., \& Matteini, L. 2018, ApJL,
  857, L19

\bibitem[{Hellinger {et~al.}(2021)Hellinger, Verdini, Landi, Franci, Papini, \&
  Matteini}]{hellal21a}
Hellinger, P., Verdini, A., Landi, S., Franci, L., Papini, E., \& Matteini, L.
  2021, Phys. Rev. Fluids, in press, arXiv:2103.12005

\bibitem[{{Kida} \& {Orszag}(1990)}]{kior90}
{Kida}, S., \& {Orszag}, S.~A. 1990, J. Sci. Comput., 5, 85

\bibitem[{Kolmogorov(1941)}]{kolm41b}
Kolmogorov, A.~N. 1941, Akademiia Nauk SSSR Doklady, 32, 16

\bibitem[{{Krupar} {et~al.}(2020){Krupar}, {Szabo}, {Maksimovic}, {Kruparova},
  {Kontar}, {Balmaceda}, {Bonnin}, {Bale}, {Pulupa}, {Malaspina}, {Bonnell},
  {Harvey}, {Goetz}, {Dudok de Wit}, {MacDowall}, {Kasper}, {Case}, {Korreck},
  {Larson}, {Livi}, {Stevens}, {Whittlesey}, \& {Hegedus}}]{krupal20}
{Krupar}, V., {et~al.} 2020, ApJS, 246, 57

\bibitem[{MacBride {et~al.}(2008)MacBride, Smith, \& Forman}]{macbal08}
MacBride, B.~T., Smith, C.~W., \& Forman, M.~A. 2008, ApJ, 679, 1644

\bibitem[{Marino {et~al.}(2008)Marino, Sorriso-Valvo, Carbone, Noullez, Bruno,
  \& Bavassano}]{marial08}
Marino, R., Sorriso-Valvo, L., Carbone, V., Noullez, A., Bruno, R., \&
  Bavassano, B. 2008, ApJL, 677, L71

\bibitem[{Matthaeus \& Velli(2011)}]{mave11}
Matthaeus, W.~H., \& Velli, M. 2011, Space Sci. Rev., 160, 145

\bibitem[{Matthaeus {et~al.}(2020)Matthaeus, Yang, Wan, Parashar,
  Bandyopadhyay, Chasapis, Pezzi, \& Valentini}]{mattal20}
Matthaeus, W.~H., Yang, Y., Wan, M., Parashar, T.~N., Bandyopadhyay, R.,
  Chasapis, A., Pezzi, O., \& Valentini, F. 2020, ApJ, 891, 101

\bibitem[{{Mininni} {et~al.}(2007){Mininni}, {Alexakis}, \&
  {Pouquet}}]{minial07}
{Mininni}, P.~D., {Alexakis}, A., \& {Pouquet}, A. 2007, J. Plasma Phys., 73,
  377

\bibitem[{{Mininni} \& {Pouquet}(2009)}]{mipo09}
{Mininni}, P.~D., \& {Pouquet}, A. 2009, Phys. Rev. E, 80, 025401

\bibitem[{Monin \& Yaglom(1975)}]{moya75}
Monin, A.~S., \& Yaglom, A.~M. 1975, Statistical fluid mechanics: Mechanics of
  turbulence (Cambridge)

\bibitem[{Montagud-Camps {et~al.}(2018)Montagud-Camps, Grappin, \&
  Verdini}]{montal18}
Montagud-Camps, V., Grappin, R., \& Verdini, A. 2018, ApJ, 853, 153

\bibitem[{Orszag(1971)}]{orsz71}
Orszag, S.~A. 1971, J. Atmospheric Sci., 28, 1074

\bibitem[{Osman {et~al.}(2011)Osman, Wan, Matthaeus, Weygand, \&
  Dasso}]{osmaal11}
Osman, K.~T., Wan, M., Matthaeus, W.~H., Weygand, J.~M., \& Dasso, S. 2011,
  Phys. Rev. Lett., 107, 165001

\bibitem[{Oughton {et~al.}(2011)Oughton, Matthaeus, Smith, Breech, \&
  Isenberg}]{oughal11}
Oughton, S., Matthaeus, W.~H., Smith, C., Breech, B., \& Isenberg, P.~A. 2011,
  J. Geophys Res., 116, A08105

\bibitem[{Oughton {et~al.}(1994)Oughton, Priest, \& Matthaeus}]{oughal94}
Oughton, S., Priest, E.~R., \& Matthaeus, W.~H. 1994, J. Fluid Mech., 280, 95

\bibitem[{Oughton {et~al.}(2013)Oughton, Wan, Servidio, \&
  Matthaeus}]{oughal13}
Oughton, S., Wan, M., Servidio, S., \& Matthaeus, W.~H. 2013, ApJ, 768, 10

\bibitem[{Papini {et~al.}(2019{\natexlab{a}})Papini, Franci, Landi, Verdini,
  Matteini, \& Hellinger}]{papial19}
Papini, E., Franci, L., Landi, S., Verdini, A., Matteini, L., \& Hellinger, P.
  2019{\natexlab{a}}, ApJ, 870, 52

\bibitem[{Papini {et~al.}(2019{\natexlab{b}})Papini, Landi, \& {Del
  Zanna}}]{papial19b}
Papini, E., Landi, S., \& {Del Zanna}, L. 2019{\natexlab{b}}, ApJ, 885, 56

\bibitem[{{Pit{\v{n}}a} {et~al.}(2019){Pit{\v{n}}a},
  {{\v{S}}afr{\'a}nkov{\'a}}, {N{\v{e}}me{\v{c}}ek}, Franci, Pi, \& {Montagud
  Camps}}]{pitnal19}
{Pit{\v{n}}a}, A., {{\v{S}}afr{\'a}nkov{\'a}}, J., {N{\v{e}}me{\v{c}}ek}, Z.,
  Franci, L., Pi, G., \& {Montagud Camps}, V. 2019, ApJ, 879, 82

\bibitem[{Podesta {et~al.}(2009)Podesta, Forman, Smith, Elton, Mal{\'e}cot, \&
  Gagne}]{podeal09}
Podesta, J.~J., Forman, M.~A., Smith, C.~W., Elton, D.~C., Mal{\'e}cot, Y., \&
  Gagne, Y. 2009, Nonlin. Proc. Geophys., 16, 99

\bibitem[{Politano \& Pouquet(1998)}]{popo98b}
Politano, H., \& Pouquet, A. 1998, Phys. Rev. E, 57, R21

\bibitem[{Praturi \& Girimaji(2019)}]{prgi19}
Praturi, D.~S., \& Girimaji, S.~S. 2019, Phys. Fluids, 31, 055114

\bibitem[{{Schmidt} \& {Grete}(2019)}]{scgr19}
{Schmidt}, W., \& {Grete}, P. 2019, Phys. Rev. E, 100, 043116

\bibitem[{{Shebalin} {et~al.}(1983){Shebalin}, {Matthaeus}, \&
  {Montgomery}}]{shebal83}
{Shebalin}, J.~V., {Matthaeus}, W.~H., \& {Montgomery}, D. 1983, J. Plasma
  Phys., 29, 525

\bibitem[{Smith {et~al.}(2009)Smith, Stawarz, Vasquez, Forman, \&
  MacBride}]{smital09}
Smith, C.~W., Stawarz, J.~E., Vasquez, B.~J., Forman, M.~A., \& MacBride, B.~T.
  2009, Phys. Rev. Lett., 103, 201101

\bibitem[{Smith \& Vasquez(2021)}]{smva21}
Smith, C.~W., \& Vasquez, B.~J. 2021, Front. Astron. Space Sci., 7, 114

\bibitem[{Smith {et~al.}(2018)Smith, Vasquez, Coburn, Forman, \&
  Stawarz}]{smital18}
Smith, C.~W., Vasquez, B.~J., Coburn, J.~T., Forman, M.~A., \& Stawarz, J.~E.
  2018, ApJ, 858, 21

\bibitem[{Sorriso-Valvo {et~al.}(2007)Sorriso-Valvo, Marino, Carbone, Noullez,
  Lepreti, Veltri, Bruno, Bavassano, \& Pietropaolo}]{sorral07}
Sorriso-Valvo, L., {et~al.} 2007, Phys. Rev. Lett., 99, 115001

\bibitem[{Stawarz {et~al.}(2009)Stawarz, Smith, Vasquez, Forman, \&
  MacBride}]{stawal09}
Stawarz, J.~E., Smith, C.~W., Vasquez, B.~J., Forman, M.~A., \& MacBride, B.~T.
  2009, ApJ, 697, 1119

\bibitem[{{\v Stver\'ak} {et~al.}(2015){\v Stver\'ak}, {Tr\'avn\'{\i}\v{c}ek},
  \& Hellinger}]{stveal15}
{\v Stver\'ak}, {\v S}., {Tr\'avn\'{\i}\v{c}ek}, P.~M., \& Hellinger, P. 2015,
  J. Geophys Res., 120, 8177

\bibitem[{Vasquez {et~al.}(2007)Vasquez, Smith, Hamilton, MacBride, \&
  Leamon}]{vasqal07}
Vasquez, B.~J., Smith, C.~W., Hamilton, K., MacBride, B.~T., \& Leamon, R.~J.
  2007, J. Geophys Res., 112, A07101

\bibitem[{Verdini {et~al.}(2015)Verdini, Grappin, Hellinger, Landi, \&
  {M\"uller}}]{verdal15}
Verdini, A., Grappin, R., Hellinger, P., Landi, S., \& {M\"uller}, W.~C. 2015,
  ApJ, 804, 119

\bibitem[{Verma(2017)}]{verm17}
Verma, M.~K. 2017, Rep. Prog. Phys., 80, 087001

\bibitem[{{Yang} {et~al.}(2017){Yang}, {Matthaeus}, {Shi}, {Wan}, \&
  {Chen}}]{yangal17b}
{Yang}, Y., {Matthaeus}, W.~H., {Shi}, Y., {Wan}, M., \& {Chen}, S. 2017, Phys.
  Fluids, 29, 035105

\bibitem[{Yang {et~al.}(2016)Yang, Shi, Wan, Matthaeus, \& Chen}]{yangal16}
Yang, Y., Shi, Y., Wan, M., Matthaeus, W.~H., \& Chen, S. 2016, Phys. Rev. E,
  93, 061102

\bibitem[{Yang {et~al.}(2019)Yang, Wan, Matthaeus, Sorriso-Valvo, Parashar, Lu,
  Shi, \& Chen}]{yangal19}
Yang, Y., Wan, M., Matthaeus, W.~H., Sorriso-Valvo, L., Parashar, T.~N., Lu,
  Q., Shi, Y., \& Chen, S. 2019, MNRAS, 482, 4933

\bibitem[{Zank {et~al.}(2017)Zank, Adhikari, Hunana, Shiota, Bruno, \&
  Telloni}]{zankal17}
Zank, G.~P., Adhikari, L., Hunana, P., Shiota, D., Bruno, R., \& Telloni, D.
  2017, ApJ, 835, 147

\bibitem[{Zhou \& Matthaeus(1990)}]{zhma90}
Zhou, Y., \& Matthaeus, W.~H. 1990, J. Geophys Res., 95, 10291

\end{thebibliography}
\end{document}